\documentclass[number,sort&compress,3p]{elsarticle}

\usepackage[numbers]{natbib}
\usepackage{graphicx}
\usepackage{float}
\usepackage[normalem]{ulem}
\usepackage{amsmath}
\usepackage[title]{appendix}
\PassOptionsToPackage{hyphens}{url}\usepackage[hidelinks]{hyperref}

\makeatletter
\def\els@aparagraph[#1]#2{\elsparagraph[#1]{#2}}
\def\els@bparagraph#1{\elsparagraph*{#1}}
\makeatother

\title{\huge Simple model of market share dynamics based on clients' firm-switching decisions}

\author[ind]{Joseph Hickey\corref{cor1}\fnref{fn1}}
\ead{joseph.hickey@alumni.ucalgary.ca}
\cortext[cor1]{Correspondence: joseph.hickey@alumni.ucalgary.ca}
\fntext[fn1]{This work was done while the author was employed at the Bank of Canada, 234 Wellington Street, Ottawa, Ontario, Canada, K1A 0G9}
\address[ind]{Independent researcher}

\begin{document}
\begin{abstract}
Firms compete for clients, creating distributions of market shares ranging from domination by a few giant companies to markets in which there are many small firms. These market structures evolve in time, and may remain stable for many years before a new firm emerges and rapidly obtains a large market share. We seek the simplest realistic model giving rise to such diverse market structures and dynamics. We focus on markets in which every client adopts a single firm, and can, from time to time, switch to a different firm. Examples include markets of cell phone and Internet service providers, and of consumer products with strong brand identification. In the model, the size of a particular firm, labelled $i$, is equal to its current number of clients, $n_i$. In every step of the simulation, a client is chosen at random, and then selects a firm from among the full set of firms with probability $p_i = (n_i^\alpha + \beta) / K$, where $K$ is the normalization factor. Our model thus has two parameters: $\alpha$ represents the degree to which firm size is an advantage ($\alpha > 1$) or disadvantage ($\alpha < 1$), relative to strict proportionality to size ($\alpha = 1$), and $\beta$ represents the degree to which small firms are viable despite their small size. We postulate that $\alpha$ and $\beta$ are determined by the regulatory, technology, business culture and social environments. The model exhibits a phase diagram in the parameter space, with different regions of behaviour. At the large $\alpha$ extreme of the phase diagram, a single dominant firm emerges, whose market share depends on the value of $\beta$. At the small $\alpha$ extreme, many firms with small market shares coexist, and no dominant firm emerges. In the intermediate region, markets are divided among a relatively small number of firms, each with sizeable market share but with distinct rankings, which can persist for long times before changing. We compare the model results to previously published empirical data from a broad range of Japanese industries, and find good agreement with a central statistical result relating the standard deviation of market share changes to the value of the market share before the change.
\end{abstract}


\maketitle

\section{Introduction}

\paragraph{} The distribution of market shares among competing firms and the stability of this distribution are fundamental aspects of a market, affecting prices, innovation, productivity, consumer satisfaction and access to services or resources, and many other economic and social factors \cite{Sutton2007a, Pike2018, Autor2020, Aghion2005, Fulton2017, Mkhaiber2021}. In light of the central importance of market structure and evolution, several basic questions arise. In a particular market, is evolution toward an absolute monopoly inevitable? Or is the market one where no clear dominant firm will ever emerge? What characteristics of a market create the potential for a new firm to enter and rapidly absorb a large fraction of the market? How often can a change in the ranking of different market participants be expected to occur, and for how long do new firms that enter the market remain active?

\paragraph{} In this paper, we propose a simple ``first-order" foundational model based on the individual movements of clients between firms. The model applies to markets in which each client has and needs a single firm, but where clients, from time to time, reconsider their choices of firm and potentially switch to a different firm. Examples of markets in which the model is intended to apply include digital economy service providers such as cell phone or Internet service providers, social media platforms, and markets of consumer goods or services in which brand choice is important.  

\paragraph{} Specifically, we consider a set of clients, each of whom is a customer of one of a set of firms. The model consists of a sequence of stochastic events. In each event, a single client is selected at random and then chooses a firm from among the full set of firms, with a certain probability. We call these events ``reconsideration events", because each event is intended to represent an individual client's process of reconsidering his or her choice of firm and potentially moving to a new firm. The probability of choosing a particular firm depends on the size (number of clients) of the firm and two parameters. The first parameter, $\alpha$, controls the degree to which having a larger size gives the firm a greater advantage in attracting clients. It represents a form of increasing ($\alpha > 1$), decreasing ($\alpha < 1$) or constant ($\alpha = 1$) returns to scale \cite{Eatwell2018}. The second parameter, $\beta$, controls a firm's probability of attracting clients regardless of its size, reflecting the ease (high values of $\beta$) or difficulty (low values of $\beta$) with which firms can enter and participate in the market. Since the client chooses a destination firm from among the full set of firms, it is possible for the client to choose the same firm he or she was with before reconsidering; in this case, the outcome represents a decision not to switch firms.

\paragraph{} The model generates market structures spanning the full range of concentration values from pure monopoly, to markets with many small-to-medium sized firms, to the theoretical low-concentration extreme in which there are as many active firms as clients and each firm has only one client. The simulated market structures can be stable in time or volatile, and include long-lived market structures with persistent firm rankings. Models of natural phenomena, including social or economic phenomena, should be constructed on a foundation that is as simple as possible in that it retains only the minimal set of model components needed to give rise to the main structural or dynamic features of the phenomenon. The fact that our model generates a broad range of complex market structures and dynamics based on only two parameters is therefore a key contribution of this work.

\paragraph{} The rest of the paper is structured as follows. Section \ref{sec:litrev} discusses the literature relevant to our model. Section \ref{sec:model} describes the model and our interpretation of the two parameters $\alpha$ and $\beta$. Section \ref{sec:results} contains simulation results, beginning, in section \ref{sec:results:alpha1_beta0}, with a special case ($\alpha=1$, $\beta=0$) in which a client's probability of selecting a destination firm is simply proportional to the size of the destination firm. Section \ref{sec:results:alpha_beta_gt_0} contains results for the more general case in which $\alpha$ and $\beta$ are positive and can range in value, and presents a phase diagram summarizing the different market structures generated by the model. Section \ref{sec:duration-leadership-and-reemerg} explores two aspects of the dynamics of the simulated market structures: the duration over which the leading firm maintains its ranking, and the lifetimes of firms that re-emerge after becoming inactive. In Section \ref{sec:comparison_empirical}, we compare the model results to previously published empirical data from Japanese industries, with discussion in section \ref{sec:discussion}.

\section{Literature review}
\label{sec:litrev}

\paragraph{} The literature on models of market structure formation and evolution is large and diverse \cite{Sutton2007a, Sutton2007b, Berry2007a, Berry2007b, Berry2021, Corchon2018, Aguirregabiria2010, Aguirregabiria2021a, Arthur1989}, and can be broadly divided into two categories: game-theoretic (both static and dynamic) and Markovian (simple stochastic) models. 

\paragraph{} In game-theoretic models, firms make strategic choices regarding factors such as production quantities, prices, investment into fixed and variable costs, and whether to enter or exit the market, in anticipation of the decisions of competing firms \cite{Berry2007a, Corchon2018, Klepper1996, Klepper2000, Klepper2006, Dunne2013}. One proposes a set of possible firm decisions and one or more economic variables to be optimized, and then seeks to determine the configuration of firm decisions in which each firm has found and made its optimal decision with respect to the anticipated decisions or responses of its competitors \cite{Sutton2007a}. The concept of strategy is thus central, and influences the structure of game-theoretic models, whether static or dynamic in form \cite{Berry2021, Aguirregabiria2021a, Aguirregabiria2021b}. Additionally, game-theoretic models are typically constructed to include economic variables such as the magnitude of set-up costs required to enter a market, how firm profits respond to entry and exit of competitors, product substitutability, entrants' expectations about competition they will face, and so on \cite{Berry2007a}. As such the models tend to be tailored to specific markets for which a high degree of data exists and for which the specific model features and strategic decision choices can be justified \cite{Sutton2007a, Berry2007a, Corchon2018}.

\paragraph{} In contrast, in Markovian models, market shares simply result from sequences of stochastic events, and the model agents (whether firms or clients) do not make strategic choices \cite{Sutton2007b, Arthur1989, Buendia2013a}. These models abstract from the details of specific industries and instead seek basic underlying processes that drive the coarse-grained behaviour of market share dynamics across a range of industries. For example, the model may simply entail a process in which each firm's market share experiences a random-walk with the number of steps in each time period being proportional to the firm's current market share \cite{Sutton2007b}. 

\paragraph{} Many Markovian models of market share dynamics belong to the family of ``urn" models, which are based on a classic statistical problem of sequentially adding balls of different colours to one or more urns \cite{Karlin1965, Mahmoud2008}. In economics, urn models have been applied to study the market structure dynamics of competing new technologies \cite{Arthur1989, Arthur1987, Dosi1994a, Dosi1994b, Bassanini2006, Marengo2017, Dosi2019, Franchini2023}, industries competing  within a geographical space \cite{Bottazzi2007} and firms' or products' market shares in international trade \cite{Dosi2015, Barbier2017, Fontanelli2023}. In such models, clients or adopters of technology are typically represented as balls, and firms or technologies are typically represented by different colours, such that adding a ball of a particular colour to the urn represents addition of a client to a certain firm or addition of an adopter of a certain technology; in this way, urn models directly incorporate the microscopic actions of individual constituents (e.g. clients).  

\paragraph{} Urn-type Markovian models that have taken a client-centred approach to market structure dynamics usually do not allow clients to change firms, but rather assume that clients are continuously added to the system over time. In such models, when a new client is added to the market, it chooses to be a customer of a particular firm or to adopt a particular technology, but does not subsequently change this choice over time \cite{Arthur1989, Buendia2013a, Buendia2013b, Weisbuch2008, Dosi2010a}. A central argument in many such models is that increasing returns to scale are needed to produce realistic market structures having features such as skewed distributions of firm sizes and path-dependent ``lock-in" (the creation of a market dominated by one firm or technology that happened by chance to obtain many adopters early in the life of the market). These models are similar to models of preferential attachment and the emergence of superstars or hits in physics and network science, in that individuals are continuously added to the system and associate with existing individuals or products, typically as a function of the existing entities' sizes or numbers of users, and the added individuals subsequently do not change their associations \cite{DeSollaPrice1976, Barabasi1999, Salganik2006}.

\paragraph{} Our model is an urn-type Markovian model, because it is based on the choices of individual clients of which firm they wish to be a customer of. However, in contrast with existing models, our model is based on the premise that individual clients ``reconsider" their choice of firm from time to time and can switch their association from their current firm to another firm.\footnote{Two recent papers investigating urn-type models allow firms or technologies to either gain or lose clients or adopters in each stochastic event. Dosi et al. \cite{Dosi2019} model adoption of one of two technologies by individual users as an urn process in which the stochastic events represent the addition of new users or the removal of existing users from either technology. The model is limited to two technologies with no possibility of entry or exit of active firms, and the only probability function examined requires a strong form of increasing returns to scale: the technology with the larger share of adopters has a larger probability of gaining a new adopter, and the technology with the smaller share of adopters has a larger probability of losing an existing adopter. Fontanelli et al. \cite{Fontanelli2023} model firm size and market share dynamics in an international trade setting with two countries using a pairwise urn process in which each stochastic event involves a competition between two firms resulting in a transfer of clients from the more productive to the less productive firm. The model in Ref. \cite{Fontanelli2023} is significantly more complex in its structure and assumptions than our model, with many more parameters and model steps, and is mainly constructed around a firm-specific feature (productivity) which is not present in our model.} This allows us to explore both the effects of gaining and of losing clients on firms' market shares, which could be a crucial model component needed to allow for dynamic evolution of a market that would otherwise lock-in to an unchanging market structure \cite{Witt1997, Bottazzi2007}.

\paragraph{} Additionally, while previous urn-type Markovian models allow for increasing returns to scale, they do not examine the effect of entry barriers. Also, more-detailed game-theoretic models typically generate market concentration via mechanisms representing a form of increasing returns to scale \cite{Ciarli2016} or entry barriers \cite{Berry2007a}, but not both. However, empirical studies have found entry barriers to be the market characteristic most clearly related to market concentration, much more so than advertising or research and development, in particular \cite{Sutton2007a, Siegfried1994, Sutton1991}. This raises fundamental questions about how existing Markovian and other urn-type or agent-based models of market structure behave when entry barriers are present. Can high market concentrations result for non-increasing returns to scale when entry barriers are present? How do market concentration and volatility behave as functions of both the degree of barrier to entry and the type and degree of return to scale? The presence of the second parameter, $\beta$, in our model allows us to explore these points.

\paragraph{} A separate branch of literature concerns models of consumer switching behaviour. Hu et al. investigated the effect of social learning on the switching choices of customers of cell phone service providers in a dynamic structural model with strategic interpersonal interactions \cite{Hu2019}. Suzuki modeled the evolution of market shares in the airline industry when clients' decisions to switch airline depend on whether they experienced a flight delay the last time they travelled \cite{Suzuki2000}. These models differ from the simple urn-type Markovian models in that they have a highly-detailed model structure involving many parameters and which is designed for application to a specific industry. In contrast, our model seeks to elucidate the general and ``first-order" underlying mechanisms driving market structure evolution, and is limited to only two control parameters. This allows us to comprehensively explore the phase space of the model across a broad range of parameter values.

\paragraph{} The most extensive dataset on market share distributions over time across many industries was studied by Sutton \cite{Sutton2007b}, and pertained to Japanese firms. Sutton showed a fundamental scaling relationship between the standard deviation of the change in market share from from period $t$ to period $t+1$ and the market share in period $t$. In section \ref{sec:comparison_empirical} we show that our results have good agreement with the scaling relationship found by Sutton.

\section{The model}
\label{sec:model} 

\paragraph{} We consider a simulation model consisting of a system of $N$ individual clients, each of whom is a customer of one of a set of firms. The firms offer a single product or service that is essential to the clients (e.g. Internet or cell phone service in a modern society) and is substitutable in the sense that the clients can satisfy the need for the good or service by switching to any other firm. To allow all possible distributions of clients among firms, including the extreme case in which there is only a single client per firm, the number of firms is equal to the number of clients, $N$. In practice, this typically means that there are many ``firms" with no clients, and these can be considered to be potential firms that could enter the market. We use the term ``active firms" to refer to firms that have, at a given point in time, at least one client.

\paragraph{} The model consists of a sequence of stochastic events, each of which represents a single client's psychological process of reconsidering his or her choice of firm and potentially switching to a new firm. Since we study the model using Monte Carlo simulations, every event corresponds to a single ``step" or ``time-step" in the simulation. In each step, a single client is randomly selected from among the population. The client then chooses a destination firm, $i$, from among the full set of firms (including the firm that the client was with upon entering the simulation step) with probability $p_i$. 

\paragraph{} In constructing the rule determining $p_i$, we first assume that a firm's probability of attracting a new client is an increasing function of its size (number of clients), $n_i$. This positive dependence on size is intended to represent the effect of increased resources that can be used to advertise, offer promotional discounts or undercut competitors, obtain prestige or recognition in society, and so on. In the simplest version of the rule, $p_i$ is directly proportional to $n_i$, such that: 
\begin{equation}
\label{eq:pi_simple}
p_i = n_i/N.
\end{equation}
This baseline model is investigated in section \ref{sec:results:alpha1_beta0}. 

\paragraph{}Building on Eq. \ref{eq:pi_simple}, we introduce two parameters. The first parameter, $\alpha$, modulates how a firm's probability of attracting clients increases with its size. The second parameter, $\beta$, allows there to be a residual probability of attracting clients, even for firms with size $n_i=0$. The probability rule becomes the following: 
\begin{equation}
\label{eq:pi}
p_i = \frac{n_i^{\alpha} + \beta}{K},
\end{equation}
where $K = \sum_{j=1}^N{n_j^\alpha} + \beta N$ is a normalization constant. As can be seen, when $\alpha=1$ and $\beta=0$, Eq. \ref{eq:pi_simple} is recovered. An increasing return to scale ($\alpha>1$) may represent markets in which clients derive benefits from being customers of firms with large customer bases, such as social media networks, online communication services including videoconferencing, and other markets where so-called ``network effects" are at play \cite{Belleflamme2018}, or where herding can occur \cite{Raafat2009, Spyrou2013}. Conversely, decreasing return to scale ($\alpha<1$), may correspond to markets in which it is difficult to offer good customer service to a larger customer base, or in which products are not significantly differentiated (such that small competitors can offer good quality and cheaper alternative products), or in which the firm's cachet or ``coolness" decreases with firm size. 

\paragraph{} The parameter $\beta$, on the other hand, provides a size-independent component to a firm's probability of attracting clients. A larger value of $\beta>0$ implies a market where it is easier for small firms to enter and participate; conversely, smaller values of $\beta>0$ correspond to higher costs of entry. The value of $\beta$ could be affected by features such as policies or technologies that allow for a baseline advertising exposure available to all firms (such as via the Internet or mandatory public registries of service providers), policies to promote competition such as a subsidies and anti-trust laws, and intellectual-property laws.

\begin{figure}[h!]
	\centering
	\includegraphics[width=0.8\textwidth]{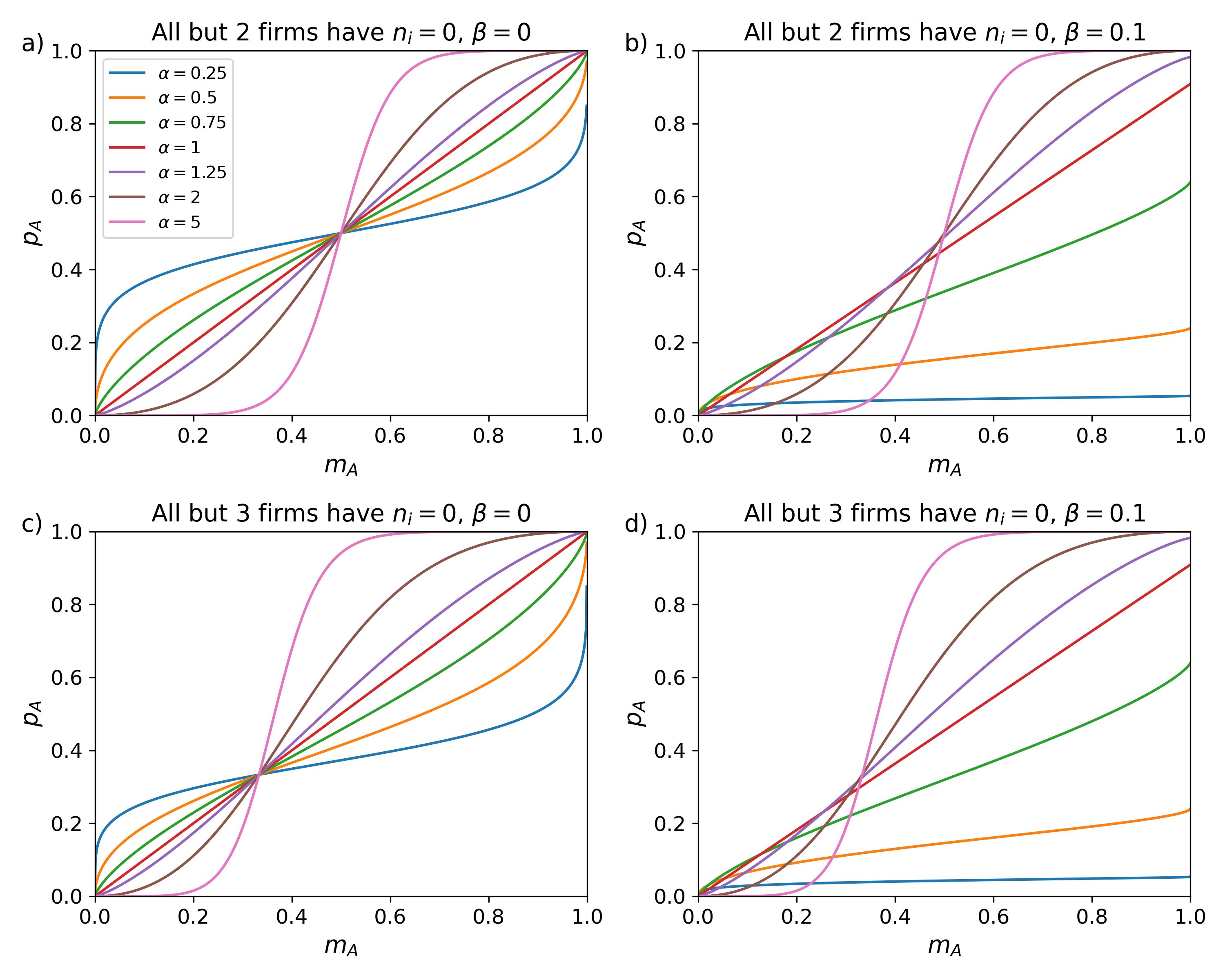}
	\caption{\small Values of $p_A$ from Eq. \ref{eq:pi}, for a particular firm (``A"), as a function of its market share, $m_A$, for two simple scenarios for the distribution of clients not currently with firm A among other firms. In (a) and (b), all clients not with firm A are with a second firm ``B", such that $m_B = 1-m_A$. In (c) and (d), all clients not with firm A are evenly split between B and a third firm ``C", such that $m_B = m_C = (1-m_A)/2$.}
	\label{fig:p_fxn}
\end{figure}

\paragraph{} Fig. \ref{fig:p_fxn} uses a few examples to help illustrate the behaviour of $p_i$ in Eq. \ref{eq:pi}. Since $p_i$ depends not only on the current size of firm $i$, but also on the current sizes of all firms other than $i$, we consider two simple scenarios for the distribution of firm sizes, for illustrative purposes.

\paragraph{} In the first scenario (Figs. \ref{fig:p_fxn}a and \ref{fig:p_fxn}b), there are two firms, called ``A" and ``B". Since there are only two firms in this scenario, all clients that are not with firm A are with firm B. The $x$-axes show all possible values of the market share of firm A, defined as $m_A = n_A/N$. The corresponding $p_A$, for various values of $\alpha$, are shown on the $y$-axes. $\beta=0$ in Fig. \ref{fig:p_fxn}a and $\beta=0.1$ in Fig. \ref{fig:p_fxn}b. 

\paragraph{} In the second scenario (Figs. \ref{fig:p_fxn}c and \ref{fig:p_fxn}d), there are three firms, ``A", ``B", and ``C", and all clients not currently with firm A are evenly split between firms B and C. As for Figs. \ref{fig:p_fxn}a and \ref{fig:p_fxn}b, we focus on firm A's probability of attracting a new client, $p_A$, as a function of $m_A$.

\paragraph{} As can be seen, $p_A$ is monotonically increasing in all cases. When $\beta=0$ (Figs. \ref{fig:p_fxn}a and \ref{fig:p_fxn}c), $p_A \to 1$ as $m_A \to 1$, for all values of $\alpha$, and the shapes of the curves depend on $\alpha$. However, when $\beta>0$ (Figs. \ref{fig:p_fxn}b and \ref{fig:p_fxn}d), $p_A$ is maximized at a value $p_A(m_A \to 1) < 1$, which decreases with decreasing $\alpha$. The intersections of the coloured curves at $m_A = 1/2$ (Fig. \ref{fig:p_fxn}a) and $m_A=1/3$ (Fig. \ref{fig:p_fxn}c) are due to the symmetries of the two scenarios considered in these examples. $p_A$ can increase sharply with increasing $m_A$, or much more gradually, depending on $\alpha$ and $\beta$. 

\paragraph{} We conclude this section by defining the relationship between ``time" in the simulation (number of simulation steps that have elapsed, $t'$) and time in the real world, $t$. The simulation entails nothing more than a sequence of events in which one of $N$ individuals is randomly chosen and reconsiders his or her choice of firm. On average, each individual client undergoes a reconsideration event once every $N$ time steps. The relationship between time and the number of simulation steps is thus:
\begin{equation}
\label{eq:t}
t =  \frac{t' \tau}{N},
\end{equation}
where $\tau$ is a constant representing the average time between occasions on which a client reconsiders his or her choice of firm in a real-world market.

\section{Simulation results}
\label{sec:results} 

\paragraph{} In this section we present simulation results showing the different market structures that emerge in the model, and how they evolve in time. We begin with the special case in which a firm's probability of attracting a client is simply proportional to the firm's current share of clients (Eq. \ref{eq:pi_simple}). This establishes several basic features of the model. In section \ref{sec:results:alpha_beta_gt_0}, we examine results for a range of values of $\alpha$ and $\beta$. Section \ref{sec:duration-leadership-and-reemerg} contains additional results regarding the statistics of leadership durations and of the lifetimes of firms that re-emerge after becoming inactive.

\subsection{Market structures when $\alpha=1$, $\beta=0$}
\label{sec:results:alpha1_beta0}

\begin{figure}[h!]
	\centering
	\includegraphics[width=\textwidth]{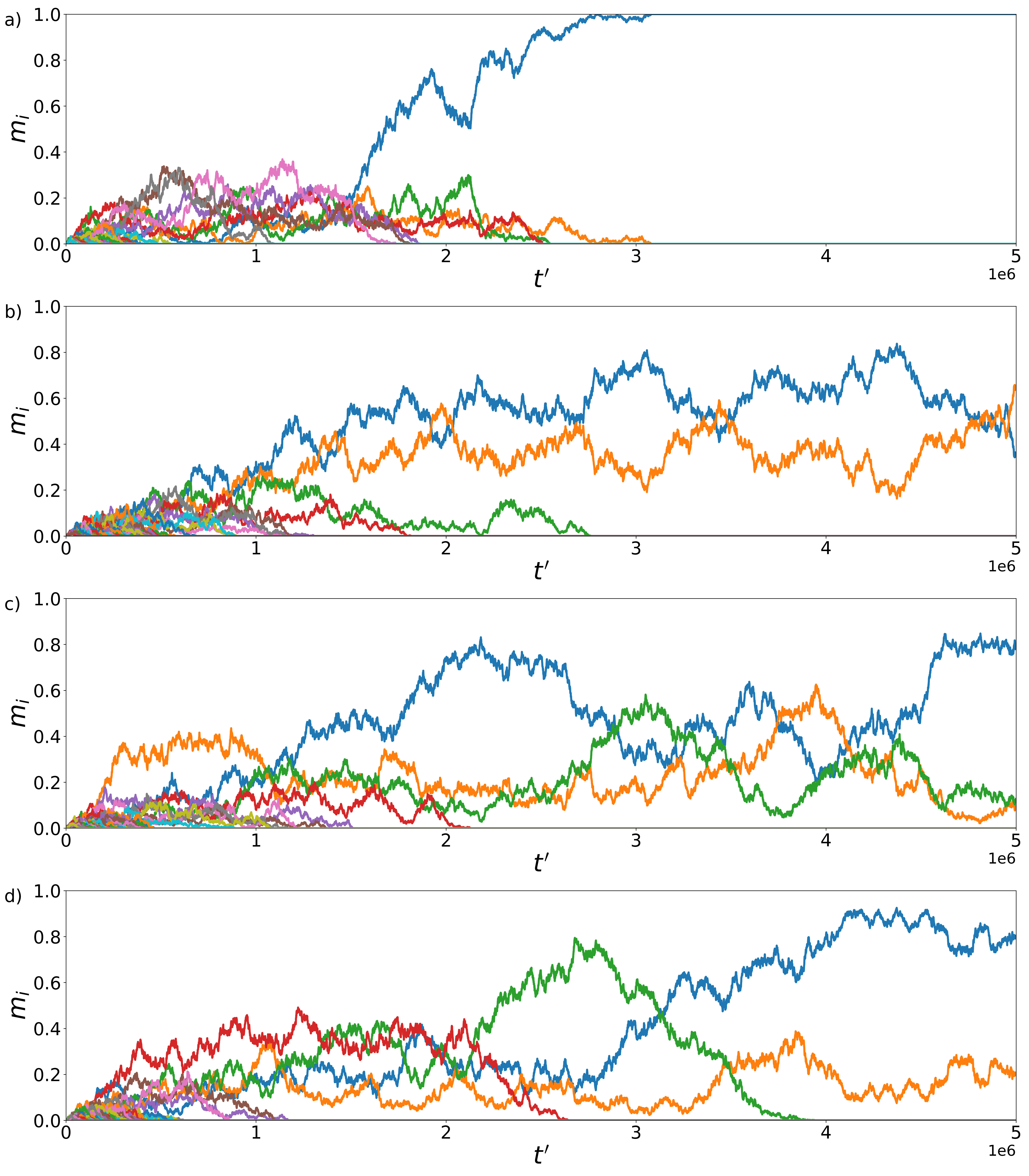}
	\caption{\small Market shares, $m_i$, of individual firms, as a function of the number of simulation time-steps, $t'$. Each of panels (a)-(d) shows a different individual realization of the simulation, beginning from an initial market share distribution of $m_i=1/N$ (i.e. $n_i=1$ for each firm $i$), for $N=2500$.}
	\label{fig:nt_alpha1_beta0}
\end{figure}

\paragraph{} Fig. \ref{fig:nt_alpha1_beta0} shows four different individual runs (``realizations") of the simulation for the special case of $\alpha=1$ and $\beta=0$. Each realization was initiated with $N=2500$ clients distributed uniformly among $N$ firms (such that each firm initially has one client) and then allowed to proceed for $T=2000\times N$ time-steps. The figure shows how the market share, $m_i = n_i/N$, of each firm evolves over time. Since each realization was performed using a different random seed of the random number generator, the results are different from one realization to the next.

\paragraph{} In this special case of the model, the system eventually arrives at a ``monopolistic" end-state in which a single firm has all of the clients. Setting $\beta=0$ in Eq. \ref{eq:pi}, one can see that if $n_i=0$, then $p_i=0$, and therefore any firm with zero clients is unable to attract any new clients and remains permanently at $n_i=0$. For a finite-sized system, random fluctuations of the $n_i$ eventually result in one firm's market share attaining $m_i=1$, at which point no further changes to the market share distribution occurs. An example can be seen in Fig. \ref{fig:nt_alpha1_beta0}a where, after a little more than $3\times 10^6$ time-steps, a single firm has obtained all clients, such that its market share is equal to 1. From that point on, the system is frozen in the monopolistic end-state.

\begin{figure}[h!]
	\centering
	\includegraphics[width=\textwidth]{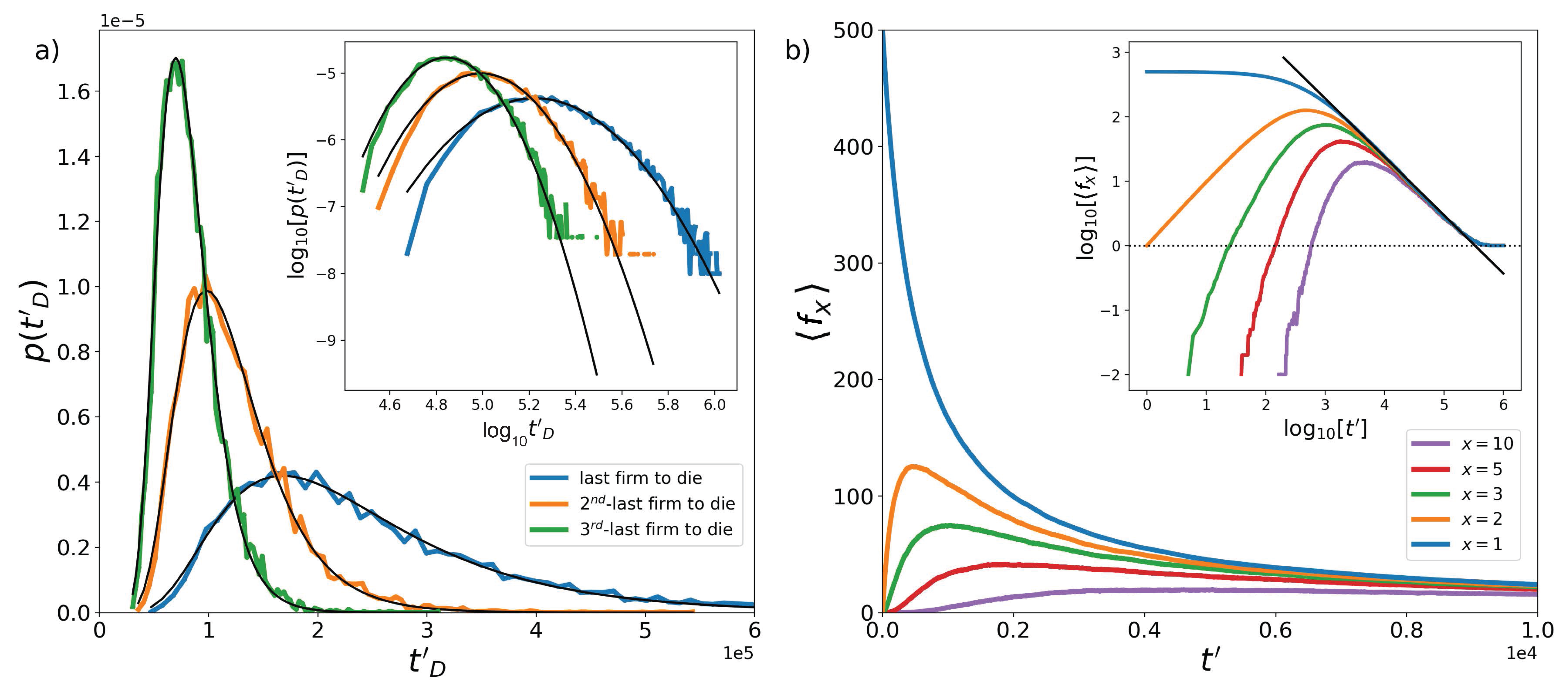}
	\caption{\small (a) Distribution of ``death times", $t'_D$, i.e., the number of simulation steps that elapse before a firm drops to $n_i=0$ clients. $t'_D$ for the ``last firm to die" (blue curve) represents the number of simulation steps that elapse before the end-state is reached in which a single firm possesses all $N$ clients. The black curves show fits of the log-normal distribution, $p(t'_D) = 1/(t'_D\sigma\sqrt{2\pi})\mathrm{exp}(-(\ln t'_D - \mu)^2 / (2\sigma^2))$. (b) The average (over many realizations) number of firms with at least $x$ clients, as a function of time. The inset shows the main plot on log-log scale. The black line in the inset is a linear fit with a slope $\approx 0.9$. A system size of $N=500$ was used and 1000 realizations of the simulation were performed in both (a) and (b).}
	\label{fig:deathtimes_alpha1_beta0}
\end{figure}

\paragraph{} However, the number of simulation steps that must elapse before the end-state is reached varies significantly from one realization to another. For example, none of the three realizations shown in Figs. \ref{fig:nt_alpha1_beta0}b-d has yet arrived at the end-state after $5\times 10^6$ simulation steps. The distribution, over many realizations, of the number of simulation steps required to reach the end-state is well-represented by a log-normal distribution (Fig. \ref{fig:deathtimes_alpha1_beta0}a).\footnote{The evolution of the market share of any single firm in our model with $\alpha=1$ and $\beta=0$ is similar to a simple one-dimensional unbiased random walk with unit step size on an interval with two absorbing barriers \cite{Redner2023}, for which the distribution of hitting times at either barrier is also well-approximated by a log-normal, although the analytic solution is more complex \cite{Jensen1970}. In our model with $\alpha=1$ and $\beta=0$, from the point of view of a given firm, each microscopic event in the model (in which a client reconsiders his or her choice of firm) can produce one of three outcomes: the firm gains a client; the firm loses a client; or there is no change to the firm's number of clients. The probability that firm $i$ gains a client is $P(\textrm{gain}) = m_i(1-m_i)$, and the probability that the firm loses a client is the same: $P(\textrm{lose}) = P(\textrm{gain})$. The dependence of $P(\textrm{gain}) = P(\textrm{lose})$ on $m_i$ is such that the probability that the firm experiences a neutral step increases as either absorbing boundary ($m_i = 0$ or $m_i = 1$) is approached, which distinguishes our model from the random walk with two absorbing barriers studied in Ref. \cite{Jensen1970}.} This means that long-lasting intermediate market structures often occur before the simulation reaches its end-state. 

\paragraph{} Fig. \ref{fig:deathtimes_alpha1_beta0}b shows the time-evolution of the number of firms that have at least $n_i = x$ clients. We use the symbol $\langle f_x \rangle$, where the angular brackets indicate an average over a set of multiple realizations of the simulation. $f_1$ represents the number of active firms, that is, the number of firms with at least one client. Fig. \ref{fig:deathtimes_alpha1_beta0}b and its inset demonstrate that the approach to the end-state is slow, going roughly like $1/t^{0.9}$. 

\paragraph{} Some of the said long-lasting intermediate market structures can be observed in Figs. \ref{fig:nt_alpha1_beta0}b-d. In Fig. \ref{fig:nt_alpha1_beta0}b, the system rapidly evolves to a configuration in which there are two clearly dominant firms (blue and orange). Beyond $t' \approx 2.75\times 10^6$, these two firms each retain about half of the clients, and one firm's gain is the other's loss. Such a configuration with two large competing firms can persist for a very long time before one of the two firms, by random chance, eventually obtains all of the clients and the size of the other firm drops to zero. In Fig. \ref{fig:nt_alpha1_beta0}c, three firms persist with non-zero market shares for many time-steps after the fourth-last-surviving (red) firm has died, and in Fig. \ref{fig:nt_alpha1_beta0}d, a four-firm market persists for many time-steps between $t' \approx 1.25 \times 10^6$ and $t' \approx 2.5 \times 10^6$. 

\paragraph{} Another important feature that can be seen in Fig. \ref{fig:nt_alpha1_beta0} is that large changes in an individual firm's market share can occur rapidly, relative to the simulation time required to reach the end-state. For example, in Fig. \ref{fig:nt_alpha1_beta0}a, the leading (blue) firm experiences a sharp drop in its market share at $t' \approx 1.9 \times 10^6$, followed by a sharp recovery at around $t' \approx 2.2 \times 10^6$. The clients lost by the blue firm in the drop are absorbed by its second-largest (green) and fourth-largest (orange) competitors, whereas the blue firm's sharp recovery comes almost entirely at the cost of the green firm. Such sharp changes in market share can also result in changes in the ranking of the firms. In Fig. \ref{fig:nt_alpha1_beta0}d, for example, beginning at around $t' \approx 2.9 \times 10^6$, the green firm is the market leader, but then undergoes two successive drops, losing clients first to the second-ranked (blue) firm, and then to the third-ranked (orange) firm, causing the green firm to quickly become the smallest of the three remaining firms, and it dies soon after.

\paragraph{} Fig. \ref{fig:nt_alpha1_beta0} is also helpful in illustrating a metric which will be used in the following section, the Herfindahl-Hirschman Index of market concentration, $HH$, which is defined as follows \cite{Rhoades1993}: 
\begin{equation}
\label{eq:HH}
HH = \sum_{i=1}^N{m_i^2} = \frac{1}{N^2}\sum_{i=1}^N{n_i^2}.
\end{equation}
\paragraph{} In our model, the concentration is maximized at $HH=1$ when one firm has all of the clients, and it is minimized at a value of $HH=1/N$, when all firms have a single client. The $HH$ values at the end of the simulations (at $t' = T = 5 \times 10^6$) in Fig. \ref{fig:nt_alpha1_beta0}a-d are 1, 0.54, 0.64, and 0.67, respectively, and at $t' = 1 \times 10^6$ the concentrations are equal to 0.16, 0.21, 0.22, and 0.24.

\subsection{Market structures when $\alpha>0$ and $\beta>0$}
\label{sec:results:alpha_beta_gt_0}

\begin{figure}[h!]
	\centering
	\includegraphics[width=\textwidth]{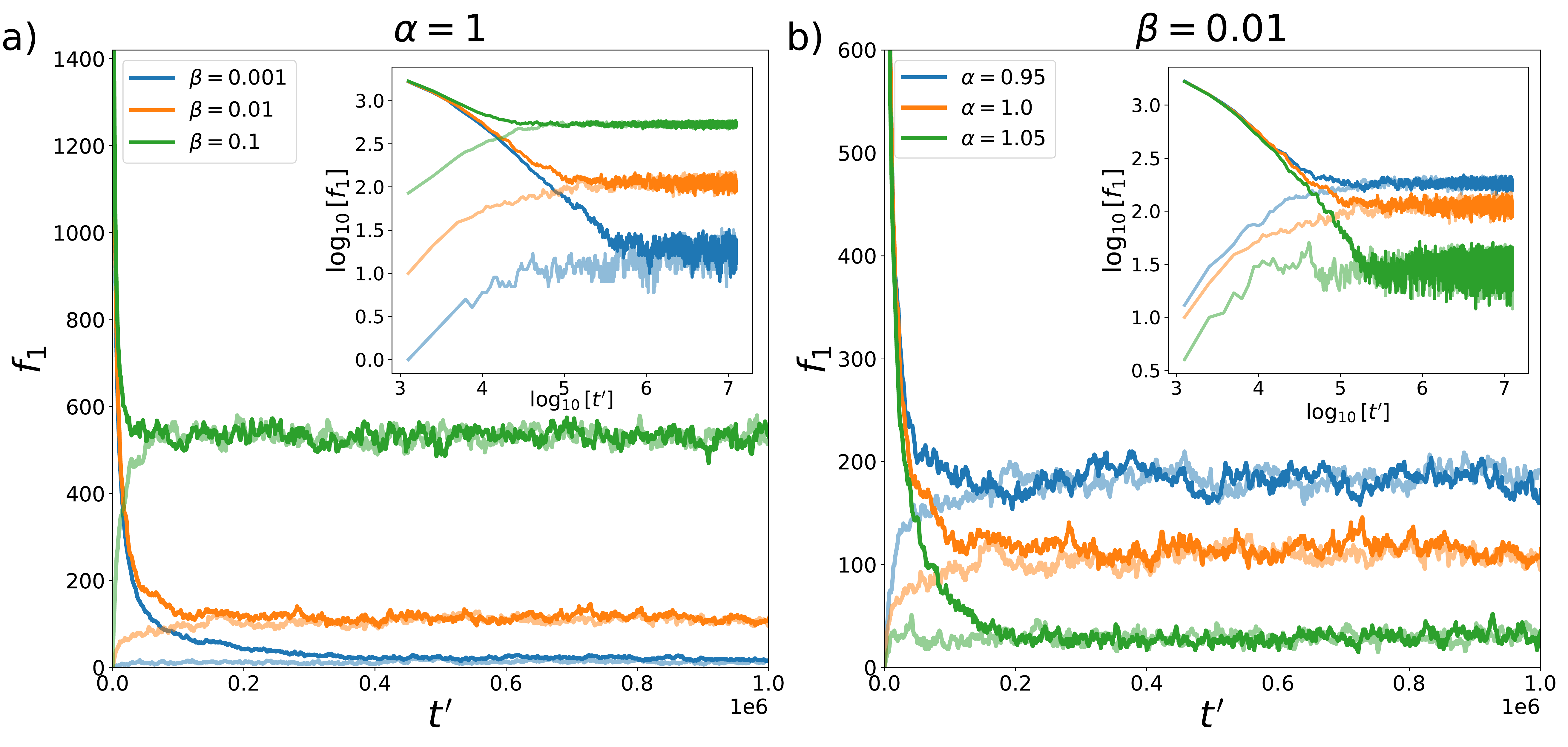}
	\caption{\small (a) Number of firms with at least 1 client, $f_1$, as a function of the number of simulation time-steps, $t'$, for $\alpha=1$ and three values of $\beta$. (b) The same quantities for fixed $\beta=0.01$ and three values of $\alpha$. Insets show the main plots on log-log scale. A system size of $N=2500$ was used. For each pair of coloured curves, the darker curve represents a single realization of the simulation beginning from the uniform initial condition in which each firm has a single client, while the lighter curve represents a single realization of the simulation beginning from the monopolostic initial condition in which a single firm has all $N$ clients. The $y$-axes of the main plots are truncated to enhance visualization.}
	\label{fig:nf_vs_t}
\end{figure}

\paragraph{} As seen in the previous section, when $\beta=0$, the system eventually arrives at an end-state in which a single firm retains all clients. The approach to this end-state was visualized in Fig. \ref{fig:deathtimes_alpha1_beta0}b, in terms of the average (over many realizations of the simulation) of $f_1$, the number firms with at least 1 client, i.e. the number of active firms. Fig. \ref{fig:nf_vs_t} explores how $f_1$ evolves for values of $\beta \geq 0$, and for various values of $\alpha$. In addition to the ``uniform" initial condition explored in section \ref{sec:results:alpha1_beta0}, in which each firm has a single client ($n_i=1$ for all firms), Fig. \ref{fig:nf_vs_t} also includes results beginning from the initial condition in which a single firm has all $N$ clients (the ``monopolistic" initial condition).

\paragraph{} Fig \ref{fig:nf_vs_t}a shows how $f_1$ evolves over time for $\alpha=1$ and for several values of $\beta$. Each pair of coloured curves in the figure shows results for the same values of $\alpha$ and $\beta$: the darker of the two curves with the same colour shows a single realization beginning from the uniform initial condition, whereas the lighter curve shows a single realization beginning from the monopolistic initial condition. For the uniform initial condition, for all choices of $\beta$, $f_1$ decreases away from its initial value and eventually arrives at a plateau value for which $f_1 > 1$. Meanwhile, for the monopolistic initial condition, for all choices of $\beta$, $f_1$ increases away from its initial value, and eventually attains the same plateau value arrived at from the uniform initial condition. This demonstrates that the system arrives at a steady-state when $\beta > 0$. Fig. \ref{fig:nf_vs_t}a shows that the steady-state number of active firms increases as $\beta$ is increased.

\paragraph{} Similarly, as shown in Fig. \ref{fig:nf_vs_t}b, fixing $\beta=0.01$ and adjusting $\alpha$ produces a steady-state value of  $f_1$. Here, we can see that larger values of $\alpha$ result in lower steady-state number of active firms.

\begin{figure}[h!]
	\centering
	\includegraphics[width=\textwidth]{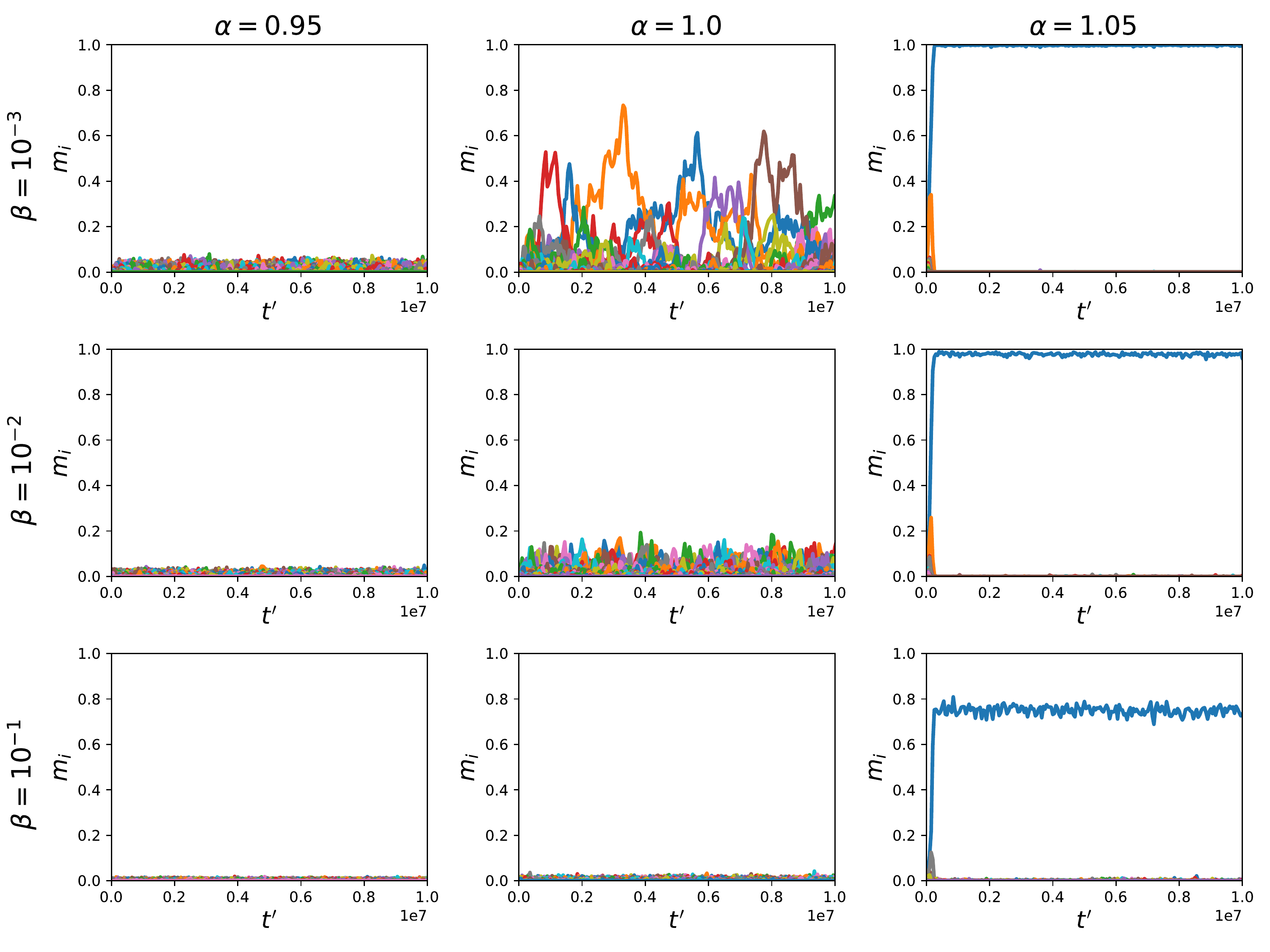}
	\caption{\small Individual realizations of the simulation for various values of $\alpha$ and $\beta>0$, showing the market shares of individual firms $m_i$ vs the number of simulation time-steps $t'$.}
	\label{fig:grid_alpha_beta_nt}
\end{figure}

\begin{figure}[h!]
	\centering
	\includegraphics[width=\textwidth]{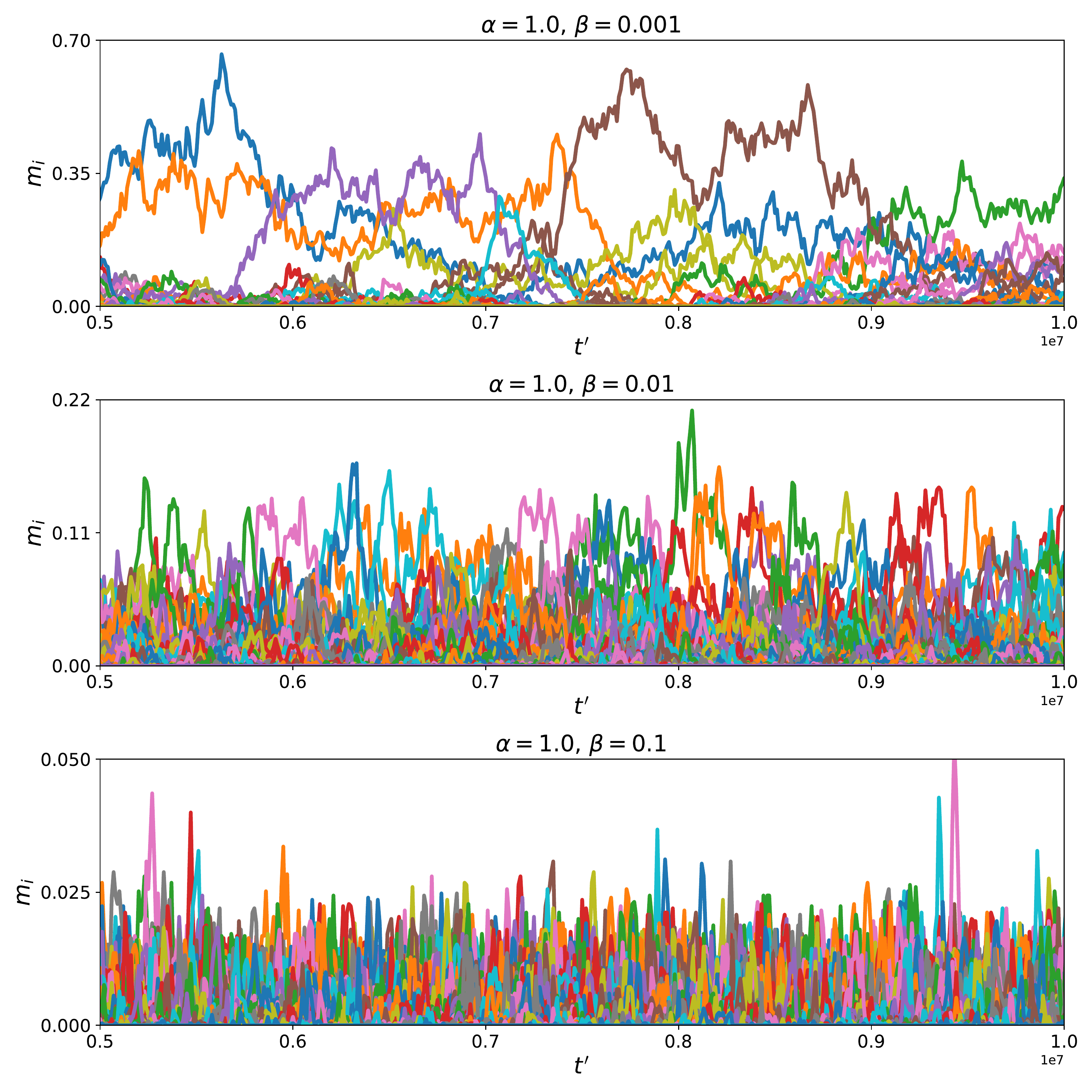}
	\caption{\small Expanded view of the panels of Fig. \ref{fig:grid_alpha_beta_nt} with $\alpha=1$, from $t' = 5 \times 10^6$ to $1 \times 10^7$.}
	\label{fig:blow_up1_from_grid}
\end{figure}

\begin{figure}[h!]
	\centering
	\includegraphics[width=\textwidth]{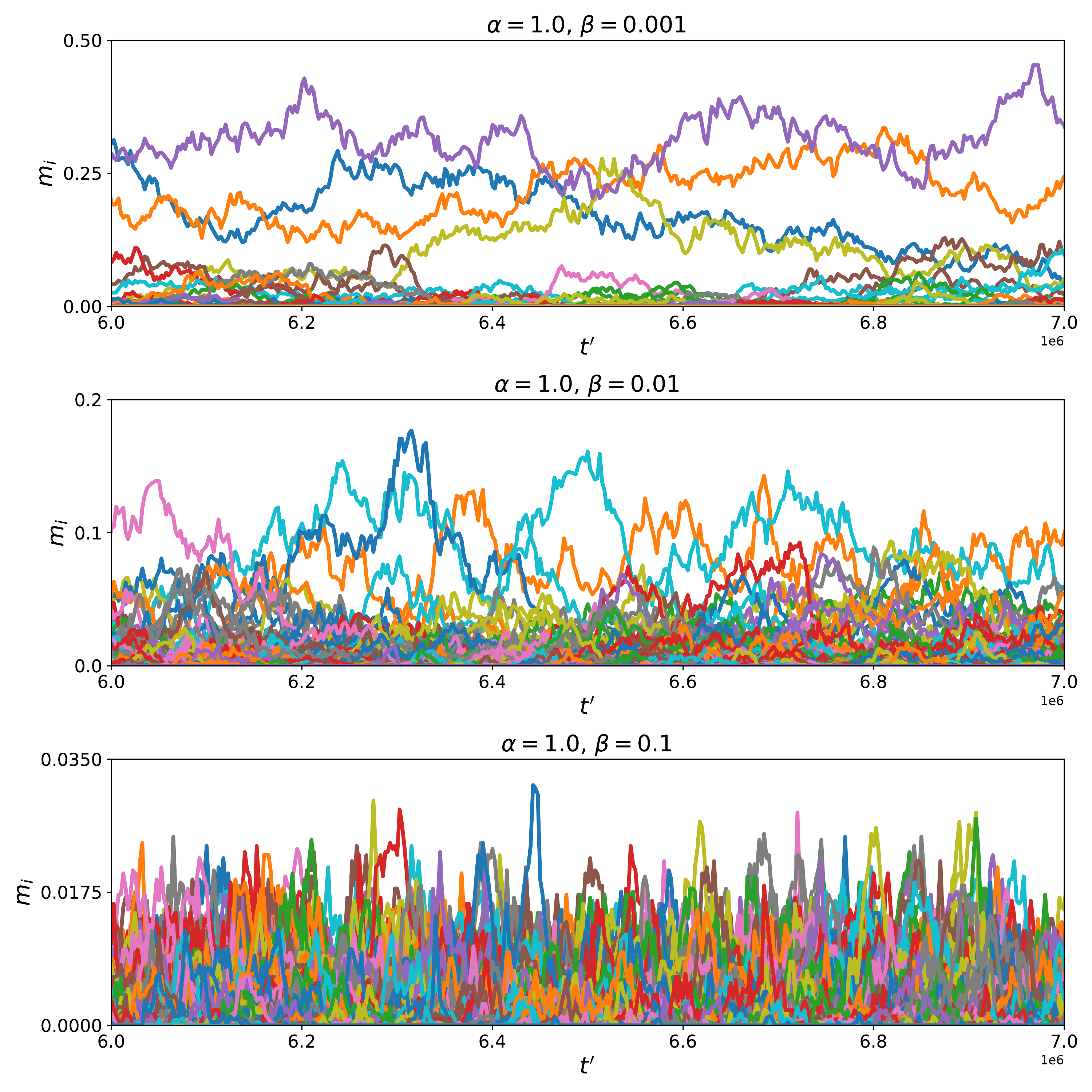}
	\caption{\small Expanded view of the panels of Fig. \ref{fig:grid_alpha_beta_nt} with $\alpha=1$, from $t' = 6 \times 10^6$ to $7 \times 10^6$.}
	\label{fig:blow_up2_from_grid}
\end{figure}

\paragraph{} Fig. \ref{fig:grid_alpha_beta_nt} shows how the market shares of individual firms evolve over time, for the values of $\alpha$ and $\beta$ used in Fig. \ref{fig:nf_vs_t}. When $\alpha=0.95$ (left column of Fig. \ref{fig:grid_alpha_beta_nt}), there are many active firms, each with a small number of clients, such that the market shares of individual firms are small. In contrast, when $\alpha=1.05$ (right column of Fig. \ref{fig:grid_alpha_beta_nt}), the market is highly concentrated, with a single firm possessing most or almost all of the clients, and the remaining clients are distributed among many small firms. There is a transition between the low-$\alpha$ and high-$\alpha$ behaviour, which can be seen in the middle column of Fig. \ref{fig:grid_alpha_beta_nt} ($\alpha=1$), which is examined more closely in Figs. \ref{fig:blow_up1_from_grid} and \ref{fig:blow_up2_from_grid}.

\newpage
\paragraph{} Figs. \ref{fig:blow_up1_from_grid} and \ref{fig:blow_up2_from_grid} show magnified views of the three $\alpha=1$ panels from Fig. \ref{fig:grid_alpha_beta_nt}. Fig. \ref{fig:blow_up1_from_grid} zooms in on the second half of the simulation time-span shown in Fig. \ref{fig:grid_alpha_beta_nt}, i.e., from $t' = 5 \times 10^6$ to $1 \times 10^7$, and Fig. \ref{fig:blow_up2_from_grid} shows a still shorter time-span, from $t' = 6 \times 10^6$ to $7 \times 10^6$. To compare with the time-scales in a real-world market, one needs to map $t'$ to $t$ via $\tau$ (Eq. \ref{eq:t}). For example, for a real-world market in which each client reconsiders his or her choice of provider once per year, on average ($\tau=1$), 10 years would be equal to $2.5 \times 10^4$ simulation time-steps in Figs. \ref{fig:grid_alpha_beta_nt}-\ref{fig:blow_up2_from_grid}. 

\paragraph{} For $\alpha=1$ and $\beta=0.001$ (top panels of Figs. \ref{fig:blow_up1_from_grid} and \ref{fig:blow_up2_from_grid}), the simulation generates a market with a relatively small number of active firms that each have a sizeable share of the $N$ clients. There are distinct rankings among the firms, and the firm rankings can persist for some time before changing, e.g. due to a leadership change in which the top-ranked firm is overtaken and replaced by a competitor. As $\beta$ is increased to $0.01$ and $0.1$, the market becomes split among more active firms with less distinct rankings. There is also a shorter persistence of the rankings; for example, the length of time over which the top-ranked firm maintains its leadership position becomes shorter as $\beta$ is increased.

\paragraph{} For $\beta = 0.1$ in Figs. \ref{fig:blow_up1_from_grid} and \ref{fig:blow_up2_from_grid}, the presence of many firms with small and relatively rapidly increasing and decreasing market shares suggests the presence of a $\beta$-dominated limit in the model, in which each time a client chooses a firm, it chooses with probability independent of firm size, such that $p_i=1/N$ for all $i$. That this is the case can be seen from an application of L'H\^opital's rule to Eq. \ref{eq:pi}, taking $\beta \to \infty$ with $\alpha$ and $N$ held constant.

\newpage
\paragraph{} To organize the simulation results into phase diagrams, we make plots of the concentration, $HH$, and the number of active firms, $f_1$, vs $\alpha$ and $\beta$, for several values of the system size, $N$. More specifically, we use the average steady-state value of the concentration $\bar{HH}$ and the average steady-state value of the number of active firms $\bar{f_1}$, obtained according to the following procedure.

\paragraph{} For each choice of $\alpha$, $\beta$ and $N$, to ensure that the simulation has arrived at its steady-state, we simultaneously run two simulations, one beginning from the uniform and one from the monopolistic initial condition. For each of the two simulations, we measure $f_1$, and record the number time-steps, $t_c'$ that elapse between $t'=0$ and the point in time at which the two simulations attain the same number of active firms. At this point in time ($t' = t_c'$), the system is deemed to have reached its steady-state plateau as shown in Fig. \ref{fig:nf_vs_t}. We then allow the simulations to proceed for a further duration equal to $5t_c'$, and record $HH$ and $f_1$ for each simulation, every $N$ time-steps. Finally, we take the time-average over the $5t_c/N$ samples to obtain $\bar{HH}$ and $\bar{f_1}$ for each of the two simulations pertaining to the two initial conditions.

\paragraph{} There is also a boundary effect which must be avoided, which is that as $\beta$ is decreased, $\bar{f_1}$ decreases, and $f_1$, which fluctuates around $\bar{f_1}$, can hit its lower limit of $f_1=1$. Therefore, as $\beta$ is decreased, to obtain valid measures of $\bar{HH}$ and $\bar{f_1}$, larger system sizes must be used in order to ensure that the simulations do not hit $f_1$ once the steady-state has been attained.

\begin{figure}[h!]
	\centering
	\includegraphics[width=\textwidth]{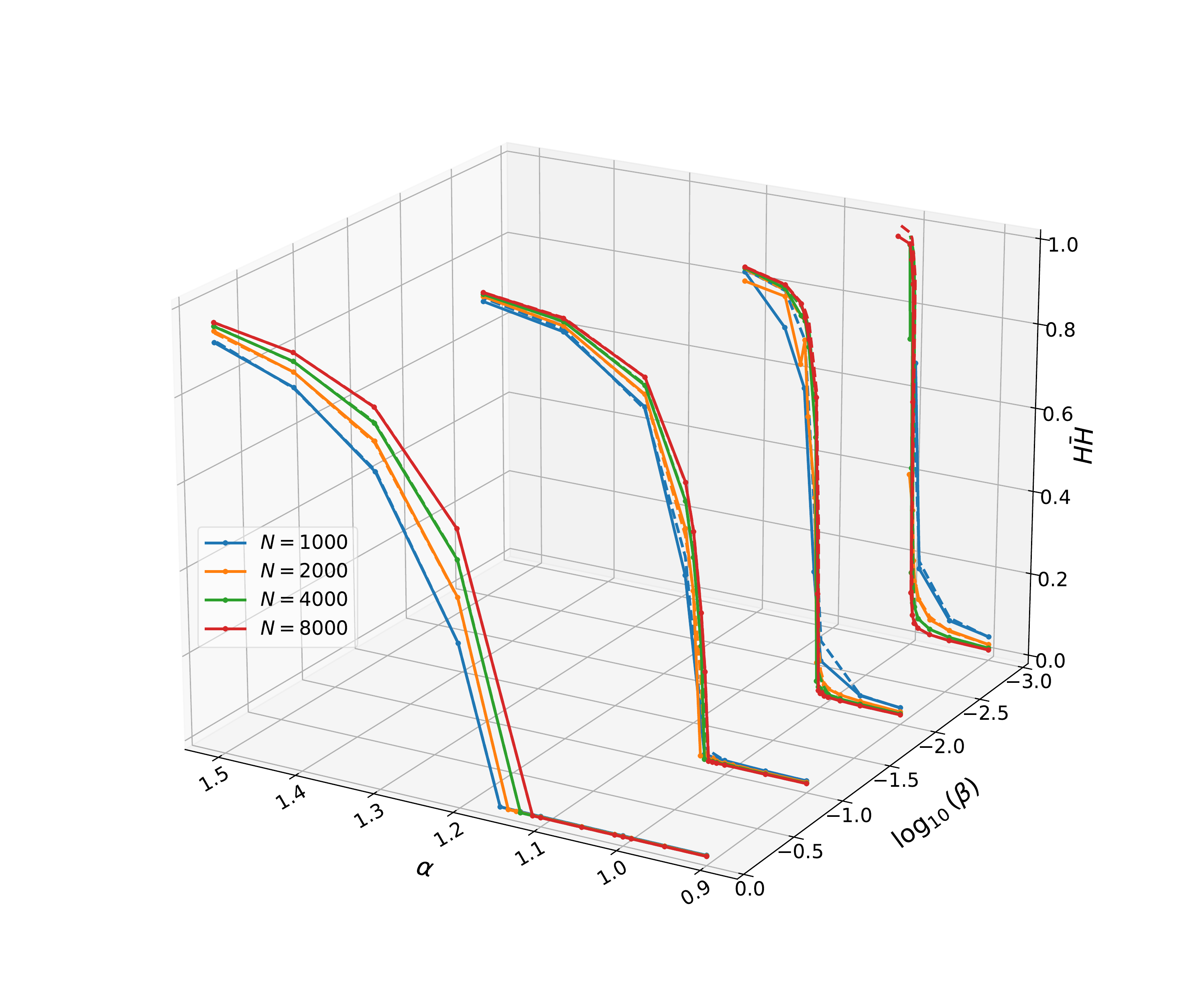}
	\caption{\small Phase diagram showing the time-averaged concentration $\bar{HH}$ as a function of $\alpha$ and $\log(\beta)$, for several values of the system size $N$. Solid lines correspond to simulations beginning from the uniform initial condition, and dashed lines correspond to simulations beginning from the monopolistic initial condition. For fixed $N$, fewer data points can be obtained as $\beta$ is decreased, due to the boundary effect discussed in the text.}
	\label{fig:pd_HH}
\end{figure}

\paragraph{} The phase diagram for the concentration $\bar{HH}$ as a function of $\alpha$ and $\log(\beta)$ is shown in Fig. \ref{fig:pd_HH}. For small values $\beta$, and for increasing $\alpha$, the concentration undergoes a transition from $\bar{HH} \approx 0$ to $\bar{HH} \approx 1$ near $\alpha=1$. The transition from low to high concentration with increasing $\alpha$ becomes more abrupt as the system size increases, reflecting a finite size effect. For larger values of $\beta$, the transition from low to high concentration occurs at larger values of $\alpha$, and the increase from $\bar{HH}$ slightly greater than 0 to 1 is more gradual than for small $\beta$, spanning a larger range of $\alpha$ values, for each system size $N$. At the same time, for fixed $\alpha>1$ and decreasing $\beta$, the concentration decreases. This can be understood by the observation, shown in the right column of Fig. \ref{fig:grid_alpha_beta_nt}, that the the market structures in this region of phase space are characterized by a single dominant firm, whose market share decreases with increasing $\beta$. As $\beta$ is decreased toward 0, the transition between low- and high-concentration regimes occurs around $\alpha=1$, such that high concentration markets do not occur when there are decreasing returns to scale ($\alpha<1$), even in the presence of high entry barriers (small $\beta$).

\begin{figure}[h!]
	\centering
	\includegraphics[width=0.8\textwidth]{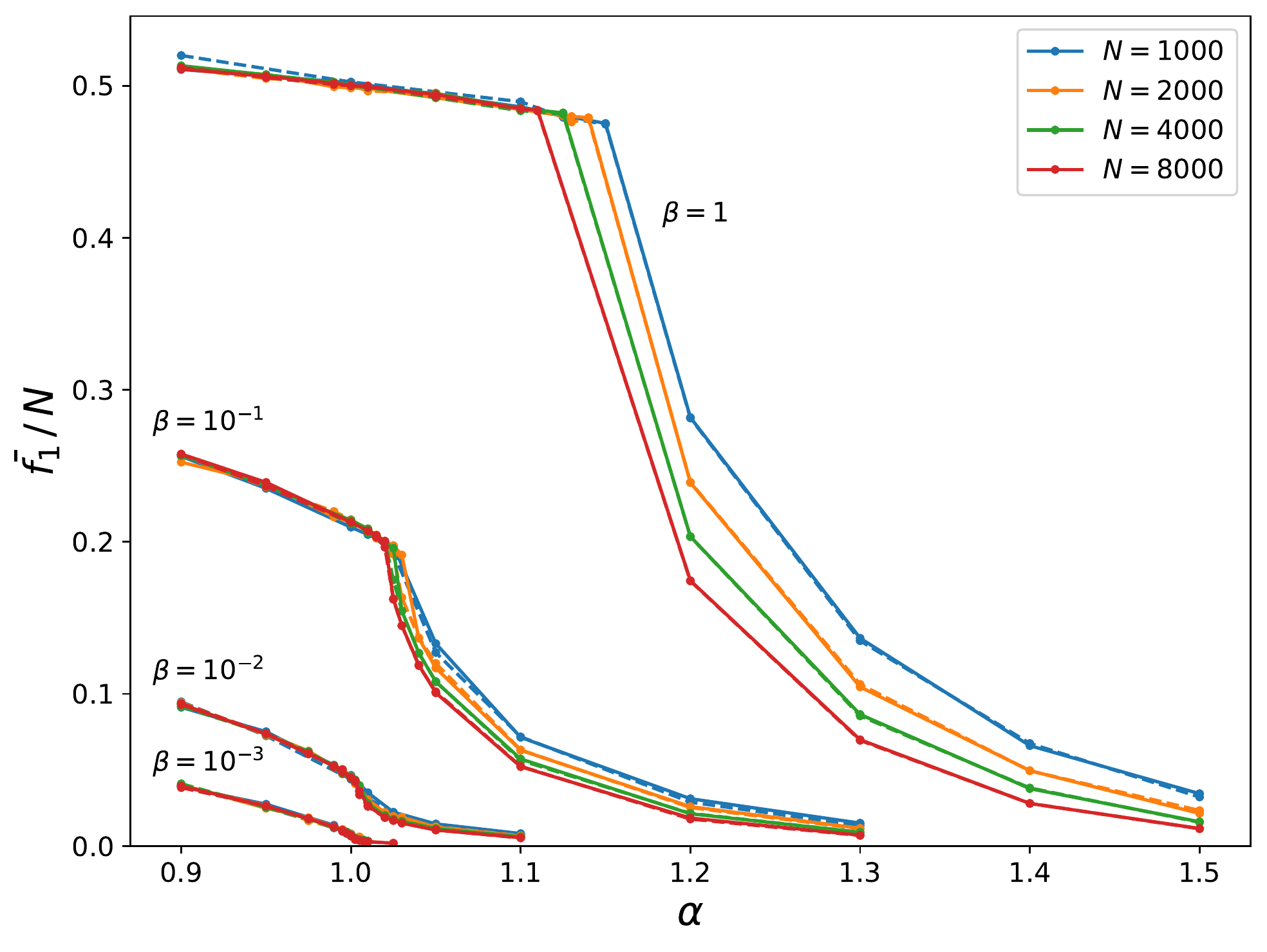}
	\caption{\small Time-averaged number of active firms scaled by system size, $\bar{f_1}/N$, versus $\alpha$, for the same values of $\beta$ and $N$ as in Fig. \ref{fig:pd_HH}. Solid lines correspond to simulations beginning from the uniform initial condition, and dashed lines correspond to simulations beginning from the monopolistic initial condition.}
	\label{fig:pd_w1}
\end{figure}

\paragraph{} Fig. \ref{fig:pd_w1} shows $\bar{f_1}/N$, the time-averaged number of active firms scaled by system size, vs $\alpha$, for the same values of $\beta$ used in Fig. \ref{fig:pd_HH}. For low values of $\alpha$, there are many active firms, and $f_1/N$ decreases slowly as $\alpha$ is increased. However, once $\alpha$ reaches the transition value, the number of active firms decreases rapidly with further increase of $\alpha$. The drop-off in the number of active firms occurs at the same value of $\alpha$ for which the concentration shoots up in Fig. \ref{fig:pd_HH}.  

\begin{figure}[h!]
	\centering
	\includegraphics[width=0.7\textwidth]{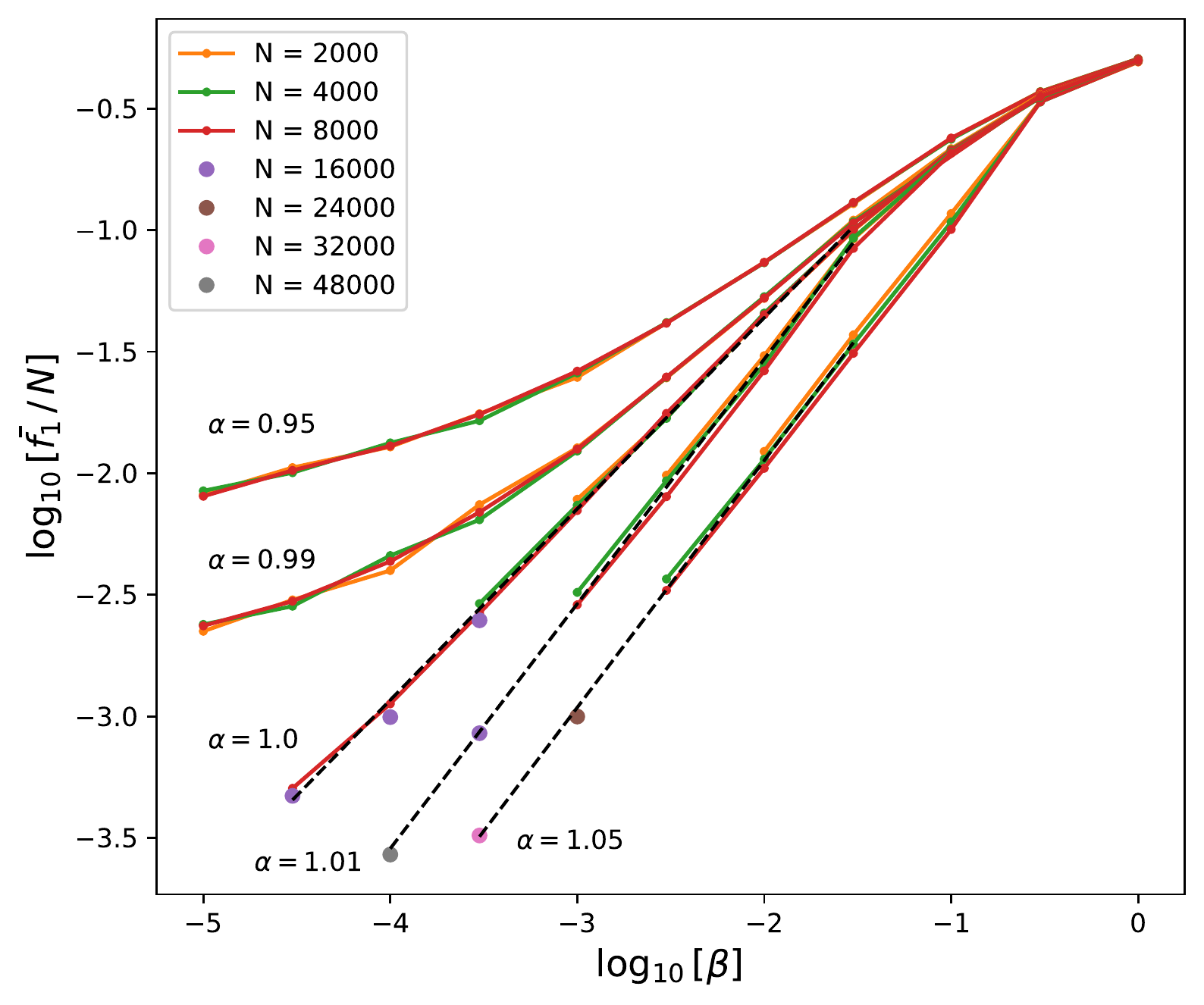}
	\caption{\small Log-log plot of the time-averaged number of active firms scaled by system size $\bar{f_1}/N$ vs $\beta$, for five different values of $\alpha$. Different system sizes are shown using different colours, as per the legend. The dashed black lines show linear fits to data for $\alpha=1$ (slope of $0.79 \pm 0.01$), $\alpha=1.01$ (slope of $1.01 \pm 0.01$) and $\alpha=1.05$ (slope of $1.02 \pm 0.02$). Each point shows the value of $\bar{f_1}/N$ resulting from a single realization of the simulation beginning from the uniform initial condition.}
	\label{fig:logf1_vs_logbeta}
\end{figure}

\newpage
\paragraph{} Fig. \ref{fig:logf1_vs_logbeta} shows how $\bar{f_1}/N$ behaves as $\beta$ is decreased to smaller values than those shown in Figs. \ref{fig:pd_HH} and \ref{fig:pd_w1}. Due to the requirement that $f_1>1$ in the simulations, which is needed to avoid the boundary effect mentioned above, Fig. \ref{fig:logf1_vs_logbeta} shows results for a number of different system sizes, with the largest system sizes (which require the longest simulation times) reserved for the smallest values of $\beta$. The steady-state was identified using the same procedure as in Figs. \ref{fig:pd_HH} and \ref{fig:pd_w1}. 

\paragraph{} As can be seen in Fig. \ref{fig:logf1_vs_logbeta}, for $\alpha<1$, the number of active firms remains relatively high and only slowly decreases as $\beta$ is decreased toward 0. This reflects the low concentration values shown in Fig. \ref{fig:pd_HH} for $\alpha<1$ and small $\beta$. For $\alpha = 1$, on the other hand, $\bar{f_1}/N$ appears to decrease as a power-law, with exponent approximately equal to $-0.8$, and for $\alpha \geq 1$, as a power-law with exponent approximately equal to $-1$. This power-law behaviour of $\bar{f_1}/N$ as $\beta \to 0$ suggests	 the presence of a phase transition occurring at $\beta=0$, for $\alpha \geq 1$ \cite{Simons1997}.

\newpage
\subsection{Duration of leadership and firm re-emergence when $\beta>0$}
\label{sec:duration-leadership-and-reemerg}

\paragraph{} In this section we examine two aspects of the dynamics of the market structures produced by the model. The first is the time-span over which the firm with the largest market share retains its leadership position ($D_L$), and the second is the lifetime of a firm that re-emerges as an active firm after previously becoming inactive due to dropping to zero clients ($D_R$). 

\begin{figure}[h!]
	\centering
	\includegraphics[width=0.65\textwidth]{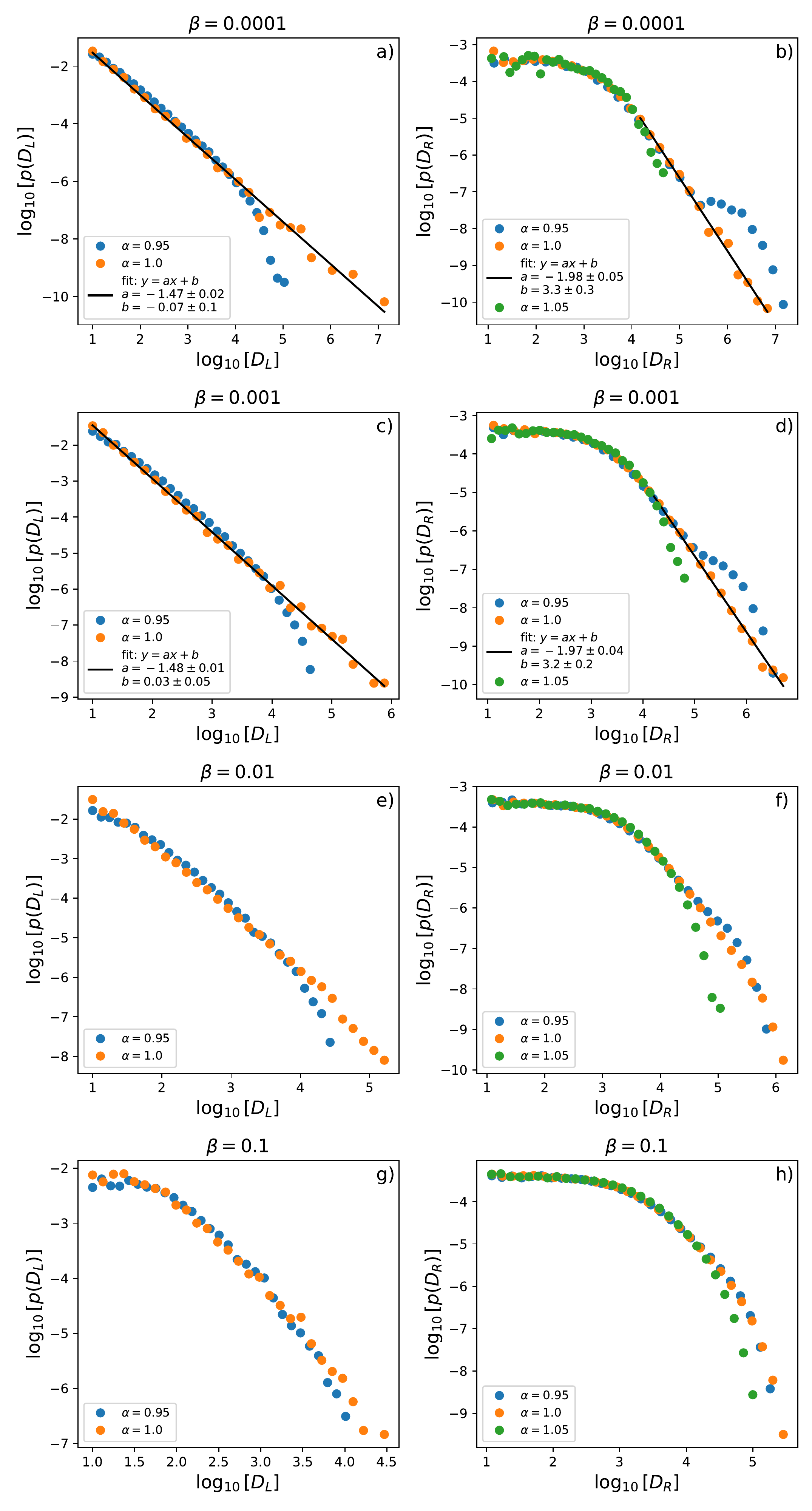}
	\caption{\small Left column of panels: Distribution of leadership duration times $D_L$ for various values of $\alpha$ and $\beta$. Right column of panels: Distibution of re-emergence lifetimes $D_R$ for various values of $\alpha$ and $\beta$. System size $N=2500$.}
	\label{fig:lct_distns_loglog}
\end{figure}

\paragraph{} Fig. \ref{fig:lct_distns_loglog} shows the distributions of $D_L$ (left column) and $D_R$ (right column) for several values of $\alpha$ and $\beta$. To create Fig. \ref{fig:lct_distns_loglog}, the same procedure as in Figs. \ref{fig:pd_HH}-\ref{fig:logf1_vs_logbeta} was used to identify the steady-state plateau, and the simulation was allowed to continue for a duration of $30t_c'$, over which time all values of $D_L$ and $D_R$ were recorded. Fig. \ref{fig:lct_distns_loglog} shows the distributions of $D_L$ and $D_R$ resulting from the simulation beginning from the uniform initial condition.

\paragraph{} Regarding $D_L$, Fig. \ref{fig:lct_distns_loglog} shows that the distribution of leadership durations follows a power-law with exponent close to $-1.5$ when $\alpha=1$ and when $\beta$ is small. The power-law extends over six orders of magnitude when $\beta=10^{-4}$ and over five orders of magnitude when $\beta=10^{-3}$. The power-law behaviour indicates that while leadership durations are often short, it is not uncommon to observe very long leadership durations, reflecting the observation in Figs. \ref{fig:blow_up1_from_grid} and \ref{fig:blow_up2_from_grid} (top panel in each figure) that firms can maintain distinct rankings for long times for $\alpha=1$ and $\beta=10^{-3}$. 

\paragraph{} For $\alpha=1$, as $\beta$ is increased to larger values ($\beta=0.01$ and $0.1$), the power-law behaviour disappears and the distribution of $D_L$ becomes similar to the distribution when $\alpha = 0.95$. This corresponds to markets with many small firms that frequently change rankings, as seen in Fig. \ref{fig:grid_alpha_beta_nt} and in the lower two panels in each of Figs. \ref{fig:blow_up1_from_grid} and \ref{fig:blow_up2_from_grid}. No data for $\alpha>1$ is included in the left column of Fig. \ref{fig:lct_distns_loglog} because there are essentially no leadership changes in the simulations, due to the presence of a single dominant firm, as can be seen in the panels of Fig. \ref{fig:grid_alpha_beta_nt} with $\alpha=1.05$.

\paragraph{} For a firm that becomes inactive by dropping to $n_i = 0$, $D_{R,i}$ is the number of simulation time-steps that elapse between the step in which the firm attains a new client (and therefore ``re-emerges", becoming active again) and the subsequent simulation step in which the firm once again drops to $n_i=0$ clients. The right column of Fig. \ref{fig:lct_distns_loglog} shows the distribution of $D_R$ for several values of $\alpha$ and $\beta$. Here, we see that the majority of firms that re-emerge have short lifetimes, for all values of $\alpha$ and $\beta$ shown in the figure. For $\alpha=1$ and small $\beta$, the tail of the distribution appears to decay as a power-law with exponent $-2$ indicating that some firms can persist for long times following re-emergence. For $\alpha = 1.05$, the tail of the distribution decays more rapidly than for $\alpha=1$, indicating the low likelihood of a long-lasting re-emergence for larger $\alpha$, while for $\alpha=0.95$, the tail of the distribution decays more slowly than for $\alpha=1$, indicating high probability of a long-lasting re-emergence for lower $\alpha$. As $\beta$ is increased to $\beta=0.01$ and $0.1$, the apparent power-law tail in the $\alpha=1$ case disappears, and the distributions for the three values of $\alpha$ become similar to one another. 

\paragraph{} Appendix \ref{sec:Appendix:Re-emerge-DR-vs-f1} contains a figure showing that there is no correlation between $f_1$ at the time of a firm's re-emergence and the subsequent lifetime, $D_R$, of the re-emergent firm.

\newpage

\newpage
\section{Comparison of model results with real-world markets}
\label{sec:comparison_empirical}

\paragraph{} In this section we compare results from the model with empirical data on the time-evolution of market shares previously published by Sutton \cite{Sutton2007b}. 

\paragraph{} Sutton analyzed a dataset consisting of annual observations of market shares in 45 different Japanese markets over 23 years. The data is of uniquely high quality in that it covers a broad range of industries, and has essentially no instances of mergers and acquisitions \cite{Sutton2007b}. From this data, Sutton found a scaling relationship between the standard deviation, $\sigma$, of the change in firms' market shares from year $t$ to $t+1$ and the market share in year $t$, equal to $\sigma = AM^{-c}$, where $A$ is a constant, $M$ is market share expressed as a percentage, and $c$ has a value slightly greater than 0.5. 

\paragraph{} Sutton obtained this result by first creating a pooled sample of the pairs of data points ($M_t$, $\Delta M_t$), where $\Delta M_t = M_{t+1} - M_t$, then binning these data pairs into equal-sized bins based on the value of $M_t$. The standard deviation of the relative change in $M_t$ for each bin, equal to $\sigma = \Delta M_t / M_t$, was then plotted vs the mean value of $M_t$ for each bin, $\bar{M_t}$. The said plot exhibited a straight line with negative slope on log-log scale, revealing the scaling relationship. The value of $c$ obtained was 0.584 (s.e. 0.053) when using 30 bins, and 0.521 (s.e. 0.024) when using 5 bins \cite{Sutton2007b}. 

\begin{figure}[h!]
	\centering
	\includegraphics[width=\textwidth]{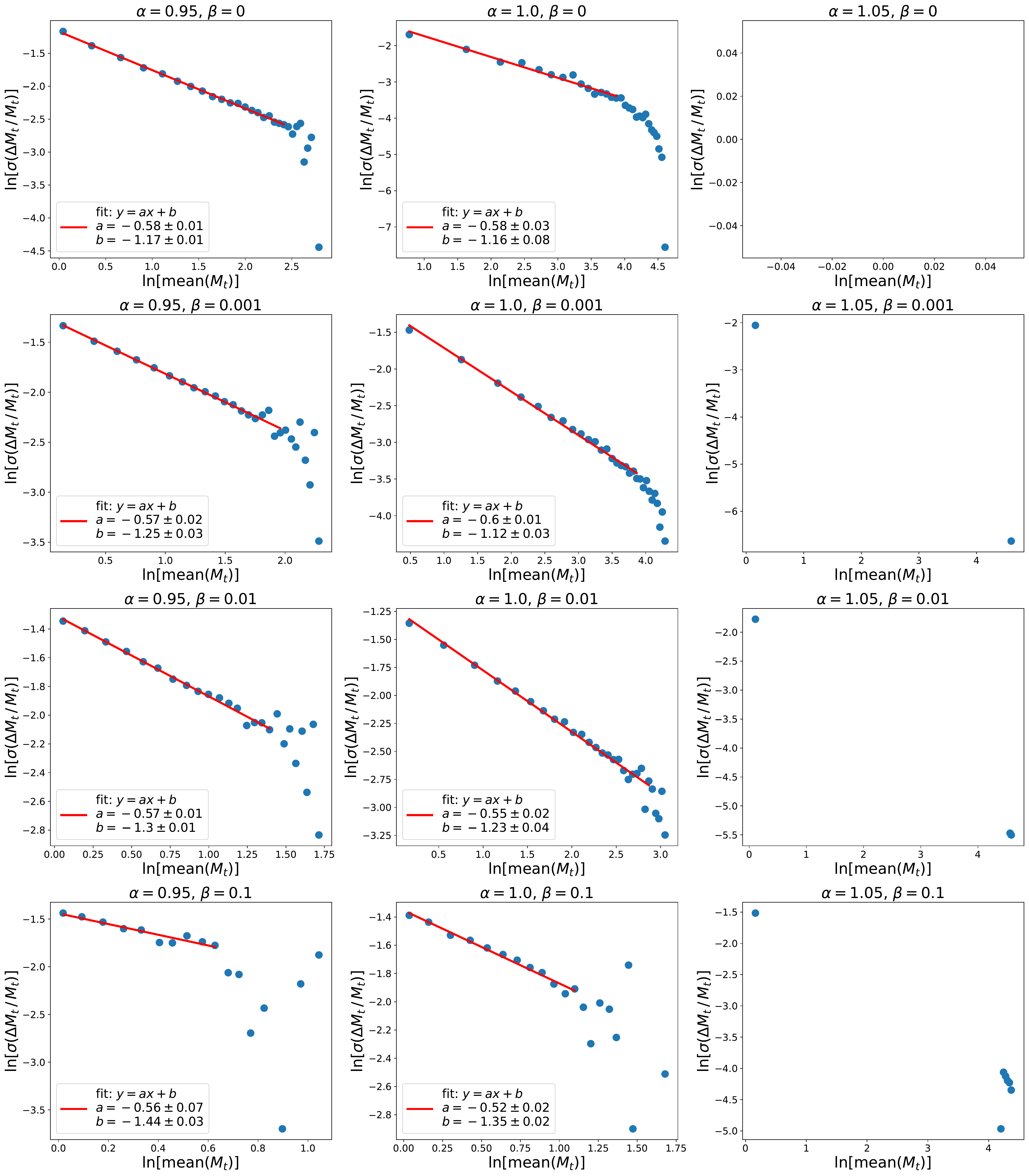}
	\caption{\small Plots of the standard deviation of the relative change in market share from the current period to the next period vs the average market share in the current period, on log-log scale. System size $N=2500$.}
	\label{fig:Sutton_loglog}
\end{figure}

\paragraph{} In Fig. \ref{fig:Sutton_loglog}, we apply the same procedure to simulated data from our model to examine how our simulation results compare with Sutton's real-world data. To do so, we observe the market shares of all firms every $N$ simulation time-steps, effectively assuming $\tau=1$ (Eq. \ref{eq:t}), and calculate the market share changes for each interval, e.g. from $t'=N$ to $t'=2N$, equivalent to the interval $t=1$ to $t=2$. The simulations in Fig. \ref{fig:Sutton_loglog} were for system size $N=2500$ and total simulation time $T=2N^2=1.25 \times 10^{7}$, using only the data from $t_1' = T/4$ to $t_2' = T$ to ensure the simulations with $\beta>0$ were in steady state (as can be seen from the insets of Fig. \ref{fig:nf_vs_t}, noting that $\log{t_1'} = 6.5$).

\paragraph{} As can be seen, for $\alpha=0.95$ (left column) and $\alpha=1$ (middle column), the simulation results have essentially the same scaling relationship found by Sutton, with the same exponent $c$ within statistical error, up to a cut-off at high-$M_t$ after which $\sigma$ decreases rapidly. In particular, the panel of Fig. \ref{fig:Sutton_loglog} with $\alpha=1.0$ and $\beta=0.001$ compares well with the results in Ref. \cite{Sutton2007b} in that the data extends up to about $\ln[\textrm{mean}(M_t)]=4$ as in Sutton's graph demonstrating the scaling relationship. This is also the parameter combination ($\alpha=1$, $\beta=0.001$) producing the most realistic concentration and ranking differentiation and persistence in Figs. \ref{fig:grid_alpha_beta_nt}-\ref{fig:blow_up2_from_grid}. For $\alpha=1.05$, since there is a single dominant firm, there is very little change in market share, and the Sutton relationship is not observed.

\newpage

\paragraph{} Sutton stated in his paper, and we show in Appendix \ref{sec:Appendix:Sutton-random-walk}, that the relationship $\sigma = AM^{-1/2}$ can result from a simple random-walk process, in which a firm's market share experiences a sequence of stochastic ``shocks" (unit increases or decreases) in each observation period (e.g. each year), with the number of shocks that occur in an observation period being proportional to the firm's market share at the beginning of the period.

\paragraph{} Unlike the simple random-walk process mentioned above, our model results (Fig. \ref{fig:Sutton_loglog}) are not consistent with the Sutton scaling relationship for large $M_t$. The mathematical reason for the different large-$M_t$ behaviour is explained in Appendix \ref{sec:Appendix:Sutton-random-walk}, and stems from the fact that our model includes the possibility of a ``neutral" event (neither gaining nor losing a client) from the point of view of the firm, which does not exist in the simple random walk process. This happens either when a client returns to his or her current firm at the conclusion of the reconsideration event, or whenever a firm is neither the current firm nor the destination firm of the client undergoing the reconsideration event.

\paragraph{} Nonetheless, despite the contrasting behaviour for large $M_t$, our model produces market share evolution that is largely consistent with the dynamics captured in the dataset of Ref. \cite{Sutton2007b}. Our model is therefore an example of a Markovian process that is based on microscopic events (individual clients' decisions about which firm to be a customer of), rather than macroscopic ``shocks" that are directly applied to the market shares of firms, that generates results that largely match a central real-world statistical observation about market structure evolution obtained from an extensive dataset. 

\section{Discussion}
\label{sec:discussion}

\paragraph{} We have presented a simple model of the formation and evolution of market structures, based on a sequence of stochastic events in which a randomly-selected individual client independently reconsiders his or her choice of firm and potentially moves to a different one. In each reconsideration event, the client selects a firm from among the full set of firms with probability depending on the size of the firm, and two parameters. The parameter $\alpha$ controls how a firm's advantage in attracting clients scales with its current number of clients. The parameter $\beta$ adds a size-independent component to a firm's probability of attracting a client, and can be interpreted as controlling the ease or difficulty with which firms can enter and participate in the market. 

\paragraph{} The model exhibits a phase diagram with different regions of behaviour (Fig. \ref{fig:pd_HH}). For small $\alpha$, many firms with small market shares coexist, and there is no dominant firm. For large values of $\alpha$, a single dominant firm emerges, and for fixed $\alpha$, the size of the dominant firm's market share decreases with increasing $\beta$. For values of $\alpha$ close to the transition point between these two regions, markets are divided among a relatively small number of firms, each with sizeable market share but with distinct rankings, which can persist for long times before changing. As $\beta$ is decreased toward 0, the transition between low- and high-concentration regimes occurs around $\alpha=1$, indicating that high concentration markets do not occur when there are decreasing returns to scale ($\alpha<1$), even in the presence of high entry barriers (small $\beta$). The long-lived dynamics of the market structures in the transition region are reflected in the apparent power-law behaviour of the duration of firms' leadership rankings and the life-times of firms that re-emerge by gaining a new client after having dropped to zero market share (Fig. \ref{fig:lct_distns_loglog}).

\paragraph{} We compare the model results to previously published empirical data from a broad range of Japanese industries, and find good agreement with a central statistical result relating the standard deviation of firms' market share changes from year $t$ to $t+1$ with the value of the market share in year $t$ (Fig. \ref{fig:Sutton_loglog}). Our model accomplishes this while being based purely on the microscopic movements of individual clients among firms and using only two parameters.

\paragraph{} Markets located toward the high-$\alpha$ region of the model's phase diagram (Fig. \ref{fig:pd_HH}) would be ones in which first-mover advantage is important due to so-called ``network effects" (i.e. that customers experience a greater benefit from being a client of a larger firm due to the firm's large user base \cite{Belleflamme2018}). Social media platforms, video-conferencing software, and ride-share apps are potential examples. Clients might derive more benefit from access to a large user base in the social media platform market, due to social connectivity with other users, than in the ride share app market, in which users typically do not interact with one another. The value of $\alpha$ might therefore be larger for social media platforms than for ride share apps. The values of $\beta$ in these markets could be related to the level of difficulty of creating and maintaining the software needed to enter and stay competitive. For a large value of $\beta$, a dominant firm can only emerge when $\alpha$ is also large (Fig. \ref{fig:pd_HH}). This suggests that, for a single dominant firm to emerge in a market that is easy to enter --- such as with relatively simple apps --- clients must be strongly attracted to or receive a lot of benefit from a large customer base, for example, by being able to connect or identify with exclusive people or celebrities associated only with the dominant firm. High-$\alpha$ markets that are dominated by a single large firm may become more competitive, with a higher fluctuation of firm rankings, if $\alpha$ is decreased into the transition region, holding $\beta$ constant.

\paragraph{} The goal of this article was to create the simplest possible model of market share structure and evolution based on the self-assignment (and self-re-assignment) of individual clients to firms, including a minimal set of parameters needed to  capture a wide range of realistic market structures and dynamics. On this basis, potential extensions can be explored. One potential area of extension concerns adding characteristics of individual firms. For example, the model could be expanded to allow two types of firms, one offering ``high-quality" and the other ``low-quality" products \cite{Dunn2008, Sweeting2020}. Another example could be the addition of a firm-specific ``loyalty" parameter that increases (or decreases) the probability with which a client remains with his or her current firm, where the strength of the loyalty parameter may depend on the time with which the client has been with the company \cite{Bolton1998}. A firm-specific $\beta$ could reflect heterogeneous \textit{ex ante} firm characteristics, which have recently been identified as important in models of firm dynamics \cite{Fontanelli2021, Sterk2021}. Individual clients may also have their own characteristics, such as one's preference to be a customer of a firm with a certain size or one's tendency to seek variety \cite{Bansal2005, Harold2020}. Switching costs might be represented in the model as a higher value of $\beta$ for the client's current firm as opposed to all other firms \cite{Farrell2007}. Finally, the baseline model presented here includes no mechanism by which firms may act. A possible area of extension would include introducing mechanisms to capture firm decisions, such as the decisions of multiple firms to merge, or competitive choices which, in a model with firm-specific versions of $\alpha$ and $\beta$, might entail a firm spending resources to adjust one of its parameters or to negatively adjust those of a competitor. A key limitation of the model presented in this article is that the system size, $N$ is fixed, whereas in real markets the number of clients (and potential firms) can increase or decrease over time. However, the model can be extended to allow for variable $N$ without changing the basic structure of the model or the probability rule in Eq. \ref{eq:pi}.

\paragraph{} Although this article focused on economic firms competing for clients, it is also relevant to many other problems. In essence, the model applies to any scenario where individual entities need to be part of or associated with a group or a system, can only belong to one group at a time, and can change their group from time to time. Scenarios in which the model could apply therefore include: political parties competing for voter shares in elections \cite{Pedersen1979}, the affiliation of individual nations to trade, military, or settlement-currency blocs \cite{Ramkishen2006, Arslanalp2022}, adoption and reformation of political systems by nations or communities \cite{Dalton2013, Lane1999}, formation of coalitions within dominance hierarchies \cite{Hickey2019, Strauss2019, Rigby2007}, gang or team membership \cite{Bolden2009, Bolden2014}, adherence to one of a set of competing social behaviours or beliefs \cite{Ditekemena2020, Ozdemir2022, Coehlo2017}, positioning in a polarized social media network \cite{Colleoni2014}, adoption of conventions or technological standards from among a set of different options \cite{Ranganathan2018, Buthe2011}, and competition for religious affiliation \cite{Leeson2017}. ``Market shares" in these diverse phenomena may be fundamentally driven by the two factors represented by the parameters $\alpha$ and $\beta$.

\newpage
\bibliographystyle{unsrtnat}
\bibliography{refs}

\newpage
\begin{appendices}
\renewcommand\thefigure{\thesection.\arabic{figure}} 
\renewcommand{\theequation}{\thesection.\arabic{equation}}  

\section{Re-emergence lifetime $D_R$ vs $f_1/N$}
\label{sec:Appendix:Re-emerge-DR-vs-f1}
\setcounter{figure}{0} 
\setcounter{equation}{0} 

\begin{figure}[h!]
	\centering
	\includegraphics[width=0.75\textwidth]{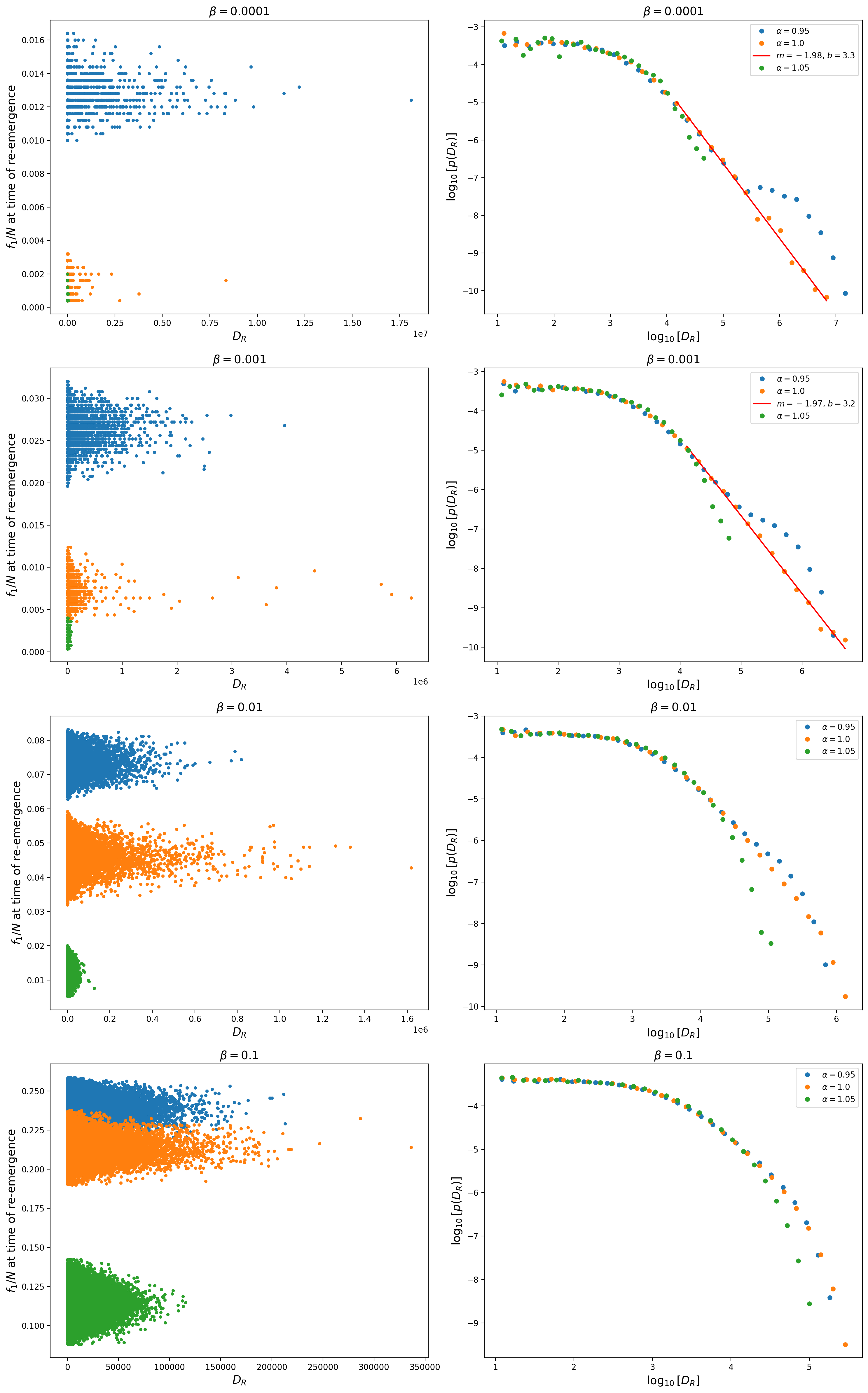}
	\caption{\small Left column of panels: $f_1/N$ for each of a set of re-emergent firms, vs the subsequent lifetime $D_R$ of the re-emergent firm. Right column of panels: Distribution of re-emergence lifetimes $D_R$ for various values of $\alpha$ and $\beta$ (same as the right column of panels of Fig. \ref{fig:lct_distns_loglog} of the main text). System size $N=2500$.}
	\label{fig:Appendix-DR-vs-f1-with-DR-distns}
\end{figure}

\paragraph{} Fig. \ref{fig:Appendix-DR-vs-f1-with-DR-distns} below examines whether a firm's re-emergence lifetime $D_R$ is related to the number of active firms in the system at the time of re-emergence. Each row of panels is for a different value of $\beta$. The left column of panels shows $f_1/N$ at the time of re-emergence vs $D_R$. The right column of panels reproduces the distributions of $D_R$ shown in Fig. \ref{fig:lct_distns_loglog} of the main text. As can be seen, from the left column of panels in Fig. \ref{fig:Appendix-DR-vs-f1-with-DR-distns}, there is no correlation between $f_1/N$ at the time of re-emergence and the subsequent lifetime $D_R$ of the re-emergent firm.

\newpage

\newpage
\section{Mathematical analysis of the scaling relationship in Section \ref{sec:comparison_empirical}}
\label{sec:Appendix:Sutton-random-walk}
\setcounter{figure}{0} 
\setcounter{equation}{0} 

\paragraph{} In the simplest definition of a one-dimensional random walk, one imagines a walker beginning at $x=0$ who then takes a sequence of discrete steps of unit size $\theta = \pm 1$, where the direction (either positive or negative) of the step is random with equal probability. After $k$ steps, the root-mean square distance (RMS) travelled by the walker is equal to $\sqrt{k}$ \cite{Tokmakoff2023}.

\paragraph{} Sutton referred to a one-dimensional random walk model in which the walker is observed at regular intervals, say every year \cite{Sutton2007b}. Within each year $t$, the walker takes $k$ steps, where $k \propto \lvert x_t\rvert$, such that the number of steps that the walker takes within a year is proportional to the absolute value of $x$ at the beginning of the year.  Since after $k$ steps, the root-mean square distance travelled by the walker is equal to $\sqrt{k}$, this means that $\textrm{RMS}(\Delta x_t) = \textrm{RMS}(x_{t+1} - x_t) =  \sqrt{k} = \sqrt{ax_t}$, where $a$ is a constant.

\paragraph{} The scaling relationship observed by Sutton (and approximated by our simulation results in Fig. \ref{fig:Sutton_loglog}) is between $\sigma(\Delta x_t / x_t)$ and $\bar{x_t}$, which are calculated after partitioning pairs of values of $\Delta x_t$ and $x_t$ into equal-sized bins by $x_t$, such that all pairs $\left(x_t, \Delta x_t\right)$ with $x_B < x_t \leq x_{B+1}$ fit into bin $B$, as in Ref. \cite{Sutton2007b}. $\sigma(\Delta x_t / x_t)$ is related to RMS($x_t$) as follows, where the sum under the square-root sign is over the $n_B$ data pairs within a bin:
\begin{equation}
\label{eq:sigma-and-RMS}
\begin{split}
\sigma\left(\frac{\Delta x_t}{x_t}\right) &= \sqrt{\frac{\sum_i^{n_B}\left[ \left( \frac{\Delta x_t}{x_t} \right)_i - \mu_B \right]^2}{n_B}} \\
&= \sqrt{\frac{\sum_i^{n_B}\left[ \left( \frac{\Delta x_t}{x_t} \right)_i \right]^2}{n_B}} \\
& \approx \frac{1}{x_t}\sqrt{\frac{\sum_i^{n_B}\left[ \left( \Delta x_t \right)_i \right]^2}{n_B}} \\
& \approx \frac{\textrm{RMS}(\Delta x_t)}{x_t} \\
& \approx \frac{\textrm{RMS}(x_{t+1} - x_t)}{x_t} \\
& \approx \frac{\sqrt{ax_t}}{x_t} \\
& \approx Ax_t^{-0.5}
\end{split}
\end{equation}
\paragraph{} The approximation in Eq. \ref{eq:sigma-and-RMS} comes from the fact that $x_t$ is approximately the same for all data pairs within a bin, and $A=\sqrt{a}$ is a constant. $\mu_B$ is the mean of $\Delta x / x_t$ for the bin $B$, and is equal to 0.

\paragraph{} We can apply the same reasoning to our model, to obtain the expected behaviour of $\sigma(\Delta x_t / x_t)$ vs $\bar{x_t}$. To do this, we think of a firm as a random walker that undergoes a number $k$ of positive or negative steps (gaining or losing a single client in each step) within an observation period consisting of $Q$ of the microscopic events in the model, in which a client is randomly selected and reconsiders his or her choice of firm.

\paragraph{} We begin with the baseline version of the model in which $\alpha=1$ and $\beta=0$. For each event in which a client is randomly selected to reconsider his or her choice of firm, one can calculate the probabilities that a particular firm with market share $m_t$ gains one client, loses one client, or has no change in its number of clients: 
\begin{equation}
\begin{split}
P(\textrm{gain}) & =  \left(1-m_t\right)m_t \\
P(\textrm{lose}) & = m_t\left(1-m_t\right) \\
P(\textrm{no change}) & = (1-m_t)(1-m_t) + m_t^2 \\
\end{split}
\end{equation}
\paragraph{} The probability, per reconsideration event, that a firm either gains or loses a client (and thus makes a step analogous to $\theta = \pm 1$ in the simple one-dimensional random walk model) is then $P(\textrm{gain or lose}) = P(\textrm{gain}) + P(\textrm{lose}) = 2m_t(1-m_t)$. Therefore, after $Q$ events have occurred, we expect a firm with market share $m_t$ to have experienced $k = 2m_t(1-m_t)Q$ non-neutral (positive or negative) ``steps" in which the firm either gained or lost one client. 

\begin{figure}[h!]
	\centering
	\includegraphics[width=0.8\textwidth]{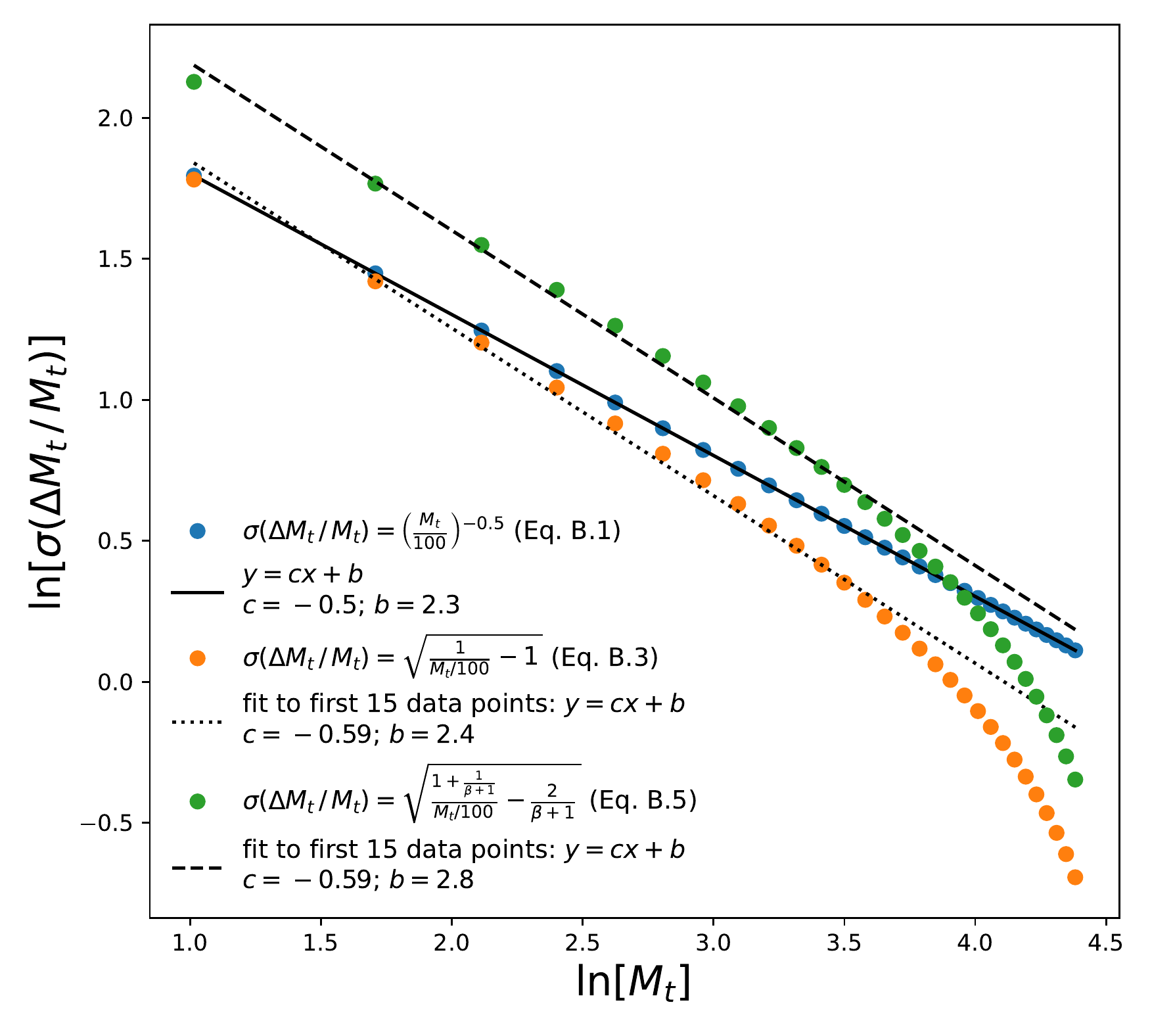}
	\caption{\small Eqs. \ref{eq:sigma-and-RMS} (blue), \ref{eq:sigma-alpha1-beta0} (orange) and \ref{eq:sigma-alpha1-beta-gt-0} (green) portrayed graphically as functions of $M_t$, for comparison with Fig. \ref{fig:Sutton_loglog} of the main text.}
	\label{fig:Appx:sigma-vs-mt-for-3-fxns}
\end{figure}

\paragraph{} Following from Eq. \ref{eq:sigma-and-RMS}, $\sigma(\Delta m_t / m_t) \approx \textrm{RMS}(\Delta m_t)/m_t$, such that $\sigma(\Delta m_t / m_t) \approx \sqrt{k}/{m_t}$, and therefore: 
\begin{equation}
\label{eq:sigma-alpha1-beta0}
\begin{split}
\sigma(\Delta m_t / m_t) & \approx \frac{\sqrt{2m_t(1-m_t)Q}}{m_t} \\
& \approx A\sqrt{\frac{m_t(1-m_t)}{m_t^2}} \\
& \approx A\sqrt{\frac{1}{m_t} - 1}
\end{split}
\end{equation}
\paragraph{} Similarly, for the case in our model in which $\alpha=1$ and $\beta > 0$, we have: 
\begin{equation}
\begin{split}
P(\textrm{gain}) & = (1-m_t)\left(\frac{n_t+\beta}{N(\beta+1)}\right) = (1 - m_t) \left( \frac{m_t}{\beta+1} + \frac{\beta}{N(\beta+1)} \right) \approx (1-m_t)\left(\frac{m_t}{\beta+1} \right) \\
P(\textrm{lose}) & = m_t\left( 1 - \frac{n_t+\beta}{N(\beta+1} \right) = m_t\left( 1 - \frac{m_t}{\beta+1} - \frac{\beta}{N(\beta+1)} \right) \approx m_t\left( 1 - \frac{m_t}{\beta+1} \right) \\
\\
\end{split}
\end{equation}
where the approximations in P(\textrm{gain}) and P(\textrm{lose}) result from taking $N \to \infty$. We then have $P(\textrm{gain or lose}) = P(\textrm{gain}) + P(\textrm{lose}) \approx m_t \left( 1 + \frac{1}{\beta+1} - \frac{2m_t}{\beta+1} \right)$. Therefore, after $Q$ events, a firm with market share $m_t$ makes $k \approx m_t \left( 1 + \frac{1}{\beta+1} - \frac{2m_t}{\beta+1} \right)Q$ ``steps" analogous to $\theta = \pm 1$ in the simple random walk model, and: 
\begin{equation}
\label{eq:sigma-alpha1-beta-gt-0}
\begin{split}
\sigma(\Delta m_t / m_t) & \approx \frac{\sqrt{k}}{m_t} \\
& \approx \frac{\sqrt{ m_t \left( 1 + \frac{1}{\beta+1} - \frac{2m_t}{\beta+1} \right)Q}}{m_t} \\
& \approx A\sqrt{\frac{1+\frac{1}{\beta+1}}{m_t} - \frac{2}{\beta+1}} \\
\end{split}
\end{equation}
\paragraph{} Fig. \ref{fig:Appx:sigma-vs-mt-for-3-fxns} shows Eqs. \ref{eq:sigma-and-RMS}, \ref{eq:sigma-alpha1-beta0} and \ref{eq:sigma-alpha1-beta-gt-0}, graphed in terms of $M_t = 100m_t$ in order to compare with Fig. \ref{fig:Sutton_loglog}. As can be seen, Eqs. \ref{eq:sigma-alpha1-beta0} and \ref{eq:sigma-alpha1-beta-gt-0} approximate the scaling relationship with exponent $-0.5$ for the main part of the data, followed by a cut-off at high-$M_t$. This is similar to the results from the simulation shown in Fig. \ref{fig:Sutton_loglog}. Furthermore, the fitted exponents (slopes in Fig. \ref{fig:Appx:sigma-vs-mt-for-3-fxns}) for Eqs. \ref{eq:sigma-alpha1-beta0} and \ref{eq:sigma-alpha1-beta-gt-0} are slightly higher than $0.5$, and very close to Sutton's result of 0.58.

\end{appendices}

\end{document}